\documentclass[useAMS,usenatbib]{mn2e}

\usepackage{lscape}
\usepackage{graphicx}
\usepackage{natbib}
\usepackage{graphics}

\title[Shapes of BCGs and normal Es]{The Shapes of BCGs and normal Ellipticals in Nearby Clusters}
\author[G. Fasano et al.]{
\parbox[t]{\textwidth}{G. Fasano$^{1}$\thanks{E-mail:giovanni.fasano@oapd.inaf.it}, D. Bettoni$^{1}$, B. Ascaso$^{2,5}$, G. Tormen$^{3}$, B.M. Poggianti$^{1}$, T. Valentinuzzi$^{3}$, M. D'Onofrio$^{3}$, J. Fritz$^{1}$, A. Moretti$^{1}$, A. Omizzolo$^{1,9}$, A. Cava$^{4,1}$, M. Moles$^{5,10}$, A. Dressler$^{6}$, W.J. Couch$^{7}$, P. Kj\ae rgaard$^{8}$ and J. Varela$^{1}$.}\\
\\
$^{1}$INAF, Osservatorio Astronomico di Padova, Vicolo Osservatorio 5, 35122 Padova, Italy\\
$^{2}$Dept. of Physics, University of California, Davis, One Shields Avenue, CA 95616, USA\\
$^{3}$Dip. Astronomia, Universit\`a di Padova, Vicolo dell'Osservatorio 2, 35122 Padova, Italy\\
$^{4}$Instituto de Astrofisica de Canarias La Laguna, Spain\\
$^{5}$Instituto de Astrofisica de Andalucia, Camino Bajo de Huetor 50, 18008 Granada, Spain\\
$^{6}$The Observatories of the Carnegie Institution of Washington, Pasadena, USA\\
$^{7}$Center for Astrophysics, Swinburne University of Technology, Australia\\
$^{8}$Copenhagen University Observatory, Niels Bohr Institute for Astronomy, Denmark\\
$^{9}$Specola Vaticana, 00120 Stato Citt\`a del Vaticano\\
$^{10}$Centro de Estudios de Fõsica del Cosmos de Aragon(CEFCA), C/General Pizarro, 1, 44001Teruel, Spain}

\begin{document}

\date{Accepted .... Received .....; in original form .....}

\pagerange{\pageref{firstpage}--\pageref{lastpage}} \pubyear{2009}

\maketitle

\label{firstpage}

\begin{abstract}

  We compare the apparent axial ratio distributions of Brightest
  Cluster Galaxies (BCGs) and normal ellipticals (Es) in our sample of
  75 galaxy clusters from the WINGS survey. Most BCGs in our clusters
  (69\%) are classified as cD galaxies. The sample of cDs has been
  completed by 14 additional cDs (non-BCGs) we found in our
  clusters. We deproject the apparent axial ratio distributions of Es,
  BCGs and cDs using a bi-variate version of the rectification Lucy's
  algorithm, whose results are supported by an independent Monte-Carlo
  technique. Finally, we compare the intrinsic shape distribution
  of BCGs to the corresponding shape distribution of the central part
  of cluster-sized dark-matter halos extracted from the GIF2
  $\Lambda$CDM $N$-body simulations.

  We find that: (i) Es have triaxial shape, the triaxiality sharing
  almost evenly the intrinsic axial ratios parameter space, with a
  weak preference for prolateness; (ii) the BCGs have triaxial shape
  as well. However, their tendence towards prolateness is much
  stronger than in the case of Es. Such a strong prolateness
  appears entirely due to 
  the sizeable (dominant) component of cDs inside the WINGS sample of BCGs.
  In fact, while the 'normal' (non-cD) BCGs do not differ from
  Es, as far as the shape distribution is concerned, the axial ratio
  distribution of BCG\_cD galaxies is found to support quite prolate
  shapes; (iii) our result turns out to be strongly at
  variance with the only similar, previous analysis by
  \citet[][RLP93]{rlp93}, where BCGs and Es were found to share the
  same axial ratio distribution; (iv) our data suggest that the above
  discrepancy is mainly caused by the different criteria that RLP93
  and ourselves use to select the cluster samples, coupled with a
  preference of cDs to reside in powerful X-ray emitting clusters; (v) the
  GIF2 $N$-body results suggest that the prolateness of the BCGs (in
  particular the cDs) could reflect the shape of the associated dark
  matter halos.

\end{abstract}


\section{Introduction} \label{secintro}

Rich galaxy clusters show in their central part a remarkable
concentration of galaxies, surrounded by progressively less dense
regions. The brightest cluster galaxies (BCGs) are usually
elliptical-like galaxies, often much brighter than the rest of the
global population of Es \citep{sanh73,sch86,sch92}.  They are normally
found close to the peaks of galaxy number density \citep{bege83} and
X-ray emission \citep{jofo84} of the cluster. The significant
alignment between the elongations of the BCGs and their host
clusters in both the optical \citep{came80,str90,plio03} and X-ray
bands \citep{has08}, together with the correlations between BCGs
luminosity and cluster properties \citep[i.e. X-ray
temperature;][]{edst91}, suggests that the formation history of BCGs
is closely linked to that of the host clusters \citep{kodj89,cozi09}.

Among the different formation scenarios which have been proposed in
the literature for BCGs, we mention: (i) merging of compact galaxy
groups before or during cluster assembly and virialization
\citep[][see also $N$-body simulations by Dubinski 1998]{ostr75,whit76,merr85}; 
(ii) filament driven, regular accretion of small, gas-rich
proto-galactic units \citep{west94}; (iii) tidal stripping from other
cluster galaxies \citep{gaos72,rich75,rich76}; (iv) continuous
accretion of star forming gas from the intracluster medium
\citep[cooling flows;][]{silk76,fabi94}; (v) late assembling ($z\sim$0.5)
of smaller, gas-poor (red) galaxies in hierarchical scenario
\citep{delu07,bern07}. Although each one of the proposed mechanisms
turns out to present some drawbacks when compared with the
observations, all of them are likely to contribute, with different
strength and timings, to the whole process of formation of BCGs.

From the observational point of view, the BCGs are special in many
aspects: besides the unusually high luminosity, the BCGs
exhibit huge sizes, sometimes further enlarged by diffuse, very
extended halos. In this case they are called cD galaxies and their
sizes may even reach $\sim$300~kpc \citep{oeml76,scho88}. The BCGs are
also extremely massive objects (up to $\sim 10^{13}M_{\odot}$), with
peculiar kinematics, in that they have lower velocity
dispersions and larger radii than predicted by the Faber-Jackson and
Kormendy relations \citep{thro81,hoes87,scho87,oeho91,bern07}, consistent
with the presence of a larger fraction of dark matter
\citep{vond07} and/or with significant growth of BCGs via dissipationless 
mergers \citep{desr07}. Finally, the BCGs often display multiple nuclei
\citep{schn83,lain03} and have been frequently identified as powerful
radio sources \citep{giac07}.

The fact that BCGs are peculiar in so many aspects with respect to the
population of normal Es, might suggest that
their shape too differs from that of Es. Also, the special role
played by BCGs in the formation history of clusters could result in a
rather peculiar shape. Normal ellipticals are
generally believed to cover the whole range of triaxiality
\citep{favi91,ryde92,bast00}, while both evolutionary scenarios supported by
simple dynamical considerations \citep{west94} and $N$-body simulations
of self-consistent models of galaxy clusters including dark matter
component \citep{dubi98}, both predict BCGs with significantly
prolate shapes. Remarkably enough, also the $N$-body simulations of
BCGs-scale, dark-matter halos in $\Lambda$CDM cosmology lead to
similar conclusions about their shape
\citep{warr92,bast05,allg06,bett07}, with the additional, interesting
hint that the dark matter halos become more prolate towards their inner 
part and at increasing the halo mass \citep{cola96,gott07}. In spite of
these converging indications, the only direct comparison available up
to now in the literature between the observed axial ratio
distributions of BCGs and Es \citep[][hereafter RLP93, observations in the
Kron--Cousins R band]{rlp93} led to
the conclusion that they are very similar.

In this paper we perform a new comparison between the distributions 
of the observed axial ratios (hereafter $q$=minor--/major--axis)
of BCGs and Es exploiting the large database of nearby
cluster galaxies provided by the WINGS survey \citep{fasa06,vare09}.
We pay particular attention to the distinction between normal BCGs and
cDs, thus including in the sample the non-BCGs, cD galaxies. We also
deproject the apparent axial ratio distributions to get the
corresponding distributions of the intrinsic axial ratios. Finally,
using the GIF2 $\Lambda$CDM cosmological simulations, we extract 3510
cluster-sized dark-matter halos and calculate the central shape at the
corresponding BCG scale, allowing us to compare their intrinsic axial
ratio distributions with those of the observed BCGs.

In Section~\ref{secsamp} we describe the galaxy sample and the axial
ratio data we use in our analysis. In Section~\ref{secaxrat} the axial
ratio distributions of Es, BCGs and cDs are presented and discussed
and the comparison with the results of RLP93 is
performed. Section~\ref{secrect} outlines the technical aspects of the
deprojection and presents the distributions of the intrinsic axial
ratios we obtain applying such deprojection techniques to the Es, BCGs
and cDs samples. In Section~\ref{secdisc} we discuss our results and
compare the intrinsic shapes of BCGs to those obtained from the GIF2
$N$-body data. Finally, Section~\ref{secsumm} summarize our
conclusions. Throughout the paper we use the
following cosmology: $H_0$=70~kms$^{-1}$Mpc$^{-1}$, $\Omega_M$=0.3 and
$\Omega_\Lambda$=0.7.

\section{Galaxy samples and axial ratio data} \label{secsamp}

We have extracted our samples of BCGs and Es from the WIde-field Nearby
Galaxy-cluster Survey \citep[WINGS;][]{fasa06}. The 77 clusters of the
WINGS sample were selected, in the redshift range 0.04-0.07, from
ROSAT catalogs of clusters
\citep{ebel96,ebel98,ebel00} and turn out to cover a wide range of
masses ($\sim 5\times 10^{13}$--$3.2\times 10^{15}~M_\odot$; 
Log$L_X\sim$43.3--44.7~ergs~s$^{-1}$). 
The optical WINGS survey provides B- and V-band imaging of the whole
sample of clusters. Integrated and aperture photometry have been
obtained for $\sim$400,000 galaxies by using SExtractor
\citep{bear96}. The procedures adopted to avoid mutual photometric
contamination between big galaxies with extended halos and smaller,
halo-embedded companions are described in detail in \citet{vare09}. We
just recall here that the largest galaxies in each cluster were
carefully modeled with IRAF-ELLIPSE and removed from the original
images in order to allow a reliable masking of the small companions
when performing the surface photometry of the big galaxies
themselves. The BCG in Abell~3164 has been excluded from the sample
since its surface photometry turned out to be uncertain due to the
proximity of the inter-chip region of the CCD mosaic, while the BGC in
Abell~3562 was not included due to the quality of the WINGS imaging 
for this cluster is not good enough to allow a reliable surface photometry
and morphology estimate. Thus, the final sample of BCGs consists of
75 galaxies.

We have classified their morphology using the WINGS V-band imaging
\citep{fasa06} and the purposely devised, automatic tool MORPHOT
\citep[Fasano et al.~2010, in preparation; see also][]{fasa07,pogg09,vale09}. 
The logical sequence and the basic procedures of MORPHOT are outlined 
in the Appendix, with particular reference to its capability of 
disentangling cDs from BCG\_Es.

On the other hand, the quantitative, multi-component analysis of the
luminosity profiles has been performed, on both the V- and B-band
WINGS images, using GASP2D \citep[Ascaso et al.~2010, in preparation; 
see also][]{mend08}. In this case, cD galaxies
have been identified by the simultaneous occurence of two conditions:
(i) the presence, in the outer profiles, of a light excess (with
respect to the inner Sersic component) significantly larger than the
photometric errors; (ii) the positioning around a surface brightness
level of $\mu_V\sim$24 of the `breaking point' where the profile
splits up from the inner profile.  When MORPHOT and GASP2D (in both V-
and B-band) produce different results or when both tools fail to
converge ($\sim$33\% of the BCG sample), we have assigned the
morphological type (E/cD) relying on the visual inspection of the
V-band images. The above procedure led us to classify most of the BCGs
in our clusters (52 galaxies) as cDs (BGC\_cD,
hereafter). Table~\ref{tab1} reports the basic information about the
BCGs in our sample, as well as some relevant properties of the host
clusters.

In the clusters Abell~3395 and Abell~3556 the BCGs are quite
off-centered with respect to the main concentration of galaxies, while
the brightest objects belonging to such concentration (in both cases
cD galaxies) turn out to be just slightly fainter than the
corresponding BCGs. This motivated us to examine the very luminous
galaxies in each cluster, searching for more non-BCG, cD candidates
(nBCG\_cD, hereafter).We found 14 additional cDs (including the two
cDs previously mentioned in the central part of Abell~3395 and
Abell~3556), all of them being just slightly fainter than the
corresponding BCGs (see Table~\ref{tab2}). Together with the 52
BCG\_cDs in Table~\ref{tab1}, these objects form a complete sample of
66 cD galaxies to be compared with the BCG\_E sample and with the
sample of normal Es extracted from the WINGS morphological catalogs
(MORPHOT). We decided to include in the sample of Es just galaxies
with absolute V magnitude $M_V<$-19.5 and cluster-centric distance
less than 0.6$R_{200}$
\footnote{$R_{200}$ is defined as the radius delimiting a sphere with
  interior mean density 200 times the critical density, approximately
  equal to the cluster virial radius.  0.6$R_{200}$ roughly
  corresponds to $R_{500}$, whose interior mean density is 500 times
  the critical density. Only a few WINGS clusters have photometric coverage
  slightly smaller than 0.6$R_{200}$ \citep{cava09}}.
The first condition guarantees that the morphological classification
is robust enough, while the second condition is necessary to homogenize
the photometric coverage of the clusters. Both conditions make the
field contamination practically negligible. After having removed the 
BCG\_Es and the cDs we are left with a final sample of 1024 Es.

In the framework of the WINGS survey, the surface photometry of
several hundreds galaxies per clusters (those with 
projected area greater than $\sim$22 arcsec$^2$ at the isophotal 
level corresponding to 2.5 times the r.m.s. of the background)
has been obtained by using
GASPHOT \citep{pign06,dono09}. This purposely devised tool performs
simultaneous best fitting of the major- and minor-axis growth light
curves of galaxies with a 2D flattened Sersic-law, convolved with the
appropriate (local) PSF. In the present paper we use for the shape
analysis the axial ratios $q_{_G}$ of the Sersic model coming out
from this best-fittig procedure. 
The reason for this choice is twofold: (i) since $q_{_G}$ is derived
from a best-fitting procedure on the whole galaxy body, it is more
stable than whatever isophotal axial ratio; (ii) since our result is
at variance with RLP93 in that we find more flattened BCGs, and since
$q_{_G}$ provides the lowest flattening among the various possible
axial ratio estimates (as shown below; see Figure~\ref{axratcomp}),
our choice is a conservative one.
Besides $q_{_G}$, GASPHOT provides the
axial ratio profiles of the galaxy isophotes. In Table~\ref{tab1} and
\ref{tab2} we report the isophotal axial ratios of the BCG and cD
galaxies corresponding to the semi major-axes of 15, 30 and 60~kpc
($q_{_{15}}$, $q_{_{30}}$ and $q_{_{60}}$). It is worth noticing that
$q_{_G}$ is not seeing-affected by definition, since it comes from the
best fit of PSF-convolved Sersic models. Instead, this is not true for
the isophotal axial ratios, at least in the innermost galaxy
regions. However, we can assume the seeing influence to vanish for
D$_{\rm m}^{\rm isoph}$/FWHM$>$3, where D$_{\rm m}^{\rm isoph}$ is the isophotal
minor-axis. In the case of our (BCG\_E$+$cD) sample, even in the worst
seeing conditions \citep[FWHM$\sim$2~arcsecs; see][]{fasa06} and for
the most flattened objects (b/a$\sim$0.3), the axial ratios of
isophotes with semi-major axis $>$10~arcsecs can be considered not
seeing-affected. For most distant WINGS clusters (z$\sim$0.07) this
translates into $\sim$13.5~kpc of isophotal semi-major axis. We
conclude that all the isophotal axial ratios reported in
Tables~\ref{tab1} and \ref{tab2} are not seeing affected. We have also
collected, for our (BCG\_E$+$cD) sample, the axial ratios $q_{_S}$
from our photometric catalogs \citep[][SExtractor]{vare09} and, if
available, the axial ratios $q_{_L}$ from the LEDA Hypercat database
\citep[][see again Tables~\ref{tab1} and \ref{tab2}]{patu03}. The last
ones are actually literature data statistically normalized to the
isophote B$_{25}$, while the first ones are luminosity weighted axial
ratios similar to those defined by RLP93, but referred to the whole
galaxy body (the axial ratios from RLP93 are instead computed within
$\sim$13~kpc). All these additional estimates of $q$ are useful to
understand how the axial ratio definition could influence our results.

\begin{figure*}
\includegraphics [scale=0.8]{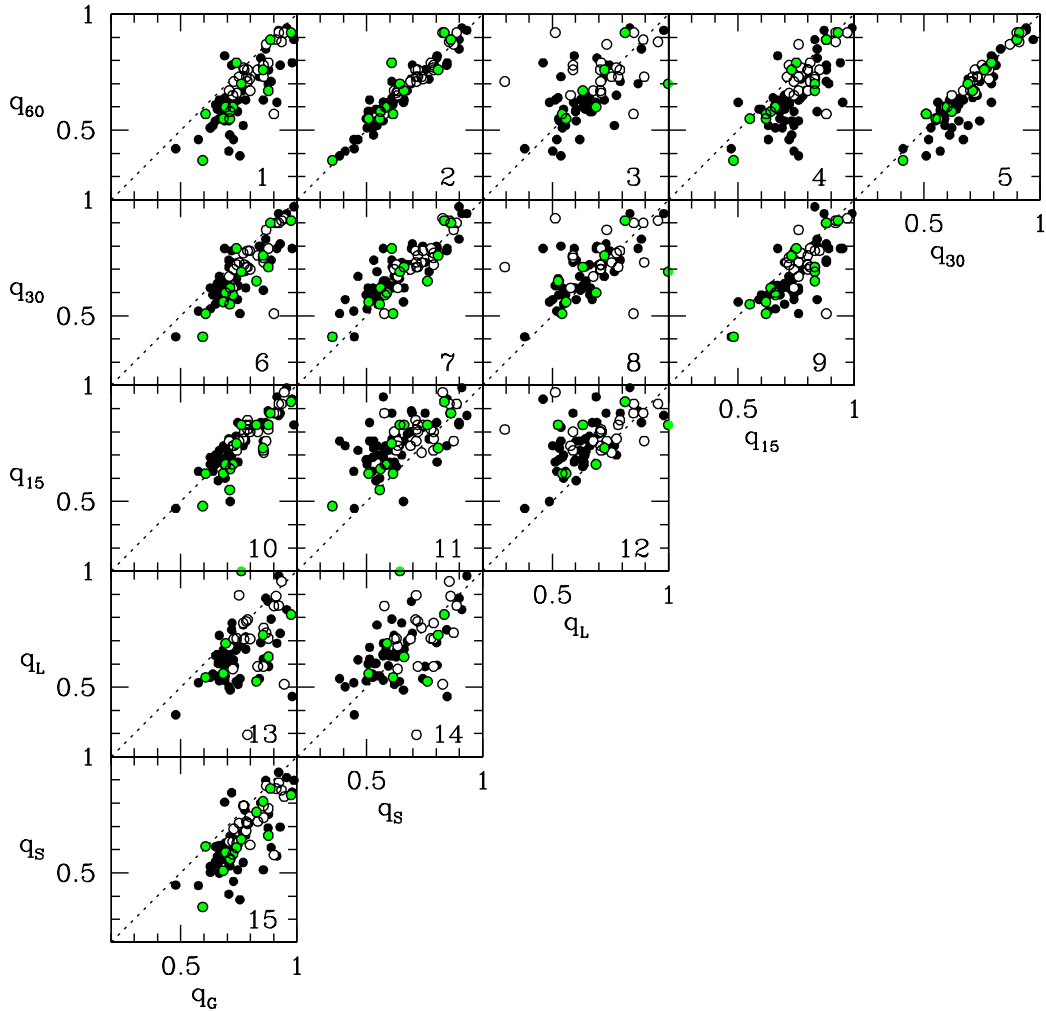}
\caption{Comparison among different axial ratio measurements for the
BCG\_Es (white dots), BCG\_cDs (black dots) and nBCG\_cDs (grey dots; green
in the electronic version) of the WINGS survey.
\label{axratcomp}}
\end{figure*}

In Figure~\ref{axratcomp} the six different estimates of the axial
ratios of our BCG\_Es$+$cDs are compared among each another. The BCG\_Es,
BCG\_cDs and nBCG\_cDs are represented in the figure by white, black
and grey (green in the electronic version) dots, respectively. 
For clarity, we will identify the panels with the numbers placed in 
bottom right. First of all, from plots 4, 5
and 9 it is clear that in our (BCG\_E$+$cD) galaxy sample the
flattening of the isophotes usually increases towards the outer galaxy
regions. From plots 13 and 15 it is also evident that the axial ratios
$q_{_G}$ are sistematically larger than both $q_{_S}$ and
$q_{_L}$. From plots 2 and 10 the axial ratios $q_{_S}$ and $q_{_G}$
turn out to be consistent with $q_{_{60}}$ and $q_{_{15}}$,
respectively. Finally, the best consistency with $q_{_L}$ is found,
although with a large scatter, in panel 8 with $q_{_{30}}$.

\section{The axial ratio distribution of BCG and normal Ellipticals} \label{secaxrat}

\begin{figure*}
\includegraphics[trim=100 0 100 0,angle=-90,scale=.65]{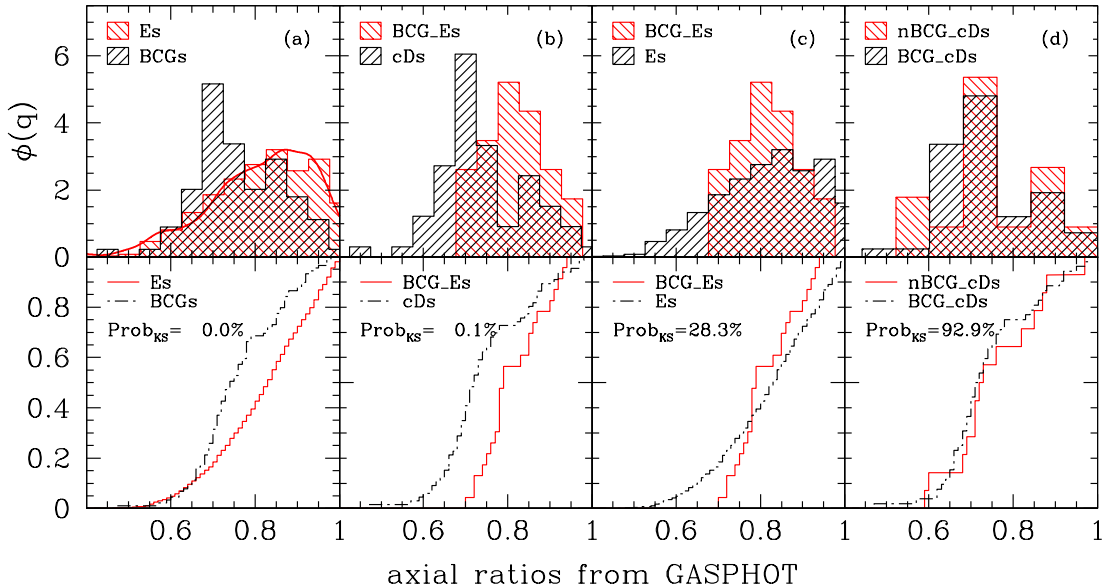}
\caption{
Differential (top panels) and cumulative (bottom panels) axial ratio
distributions in the WINGS survey: comparisons between Es and BCGs
(leftmost panels; red and black, respectively), BCG\_Es and cD galaxies
(panel~b), BCG\_Es and Es (panel~c), nBCG\_cDs and BCG\_cDs (rightmost
panels). The bottom panels also report the probability that they are drawn from the
same parent populations, according to the KS statistics.\label{axratdistr}}
\end{figure*}

In Figure~\ref{axratdistr} we compare the axial ratio distributions
$\phi (q_{_G})$ of Es and BCGs (panel~a), BCG\_Es
and cDs (panel~b), BCG\_Es and Es (panel~c), nBCG\_cDs and
BCG\_cDs (panel~d). The full line (red in the electronic version of the
paper) in the top-leftmost panel of
the figure, reports the axial ratio distribution of Es obtained by
\citet{favi91}, which turns out to be fairly in agreement with
that of WINGS Es. The bottom panels show
the corresponding cumulative distributions and report, 
in each case, the
probability of the ``null hypothesis'' (i.e. that they are drawn from the
same parent populations), according to the Kolmogorov-Smirnov (KS)
statistics. 
Figure~\ref{axratdistr} (panel~a) shows that the distribution $\phi
(q_{_G})$ of BCGs looks strongly different from that of normal ellipticals,
the BCGs being sistematically flatter than Es. 
\footnote{Note that our sample of Es includes galaxies with absolute 
V magnitude brighter than -19.5: a rather faint cutoff, indeed. Since 
luminous Es ($M_B<$-20, i.e. roughly $M_V<$-21) are found to be rounder 
\citep{trem96}, had we adopted a brighter cutoff, the difference between 
the axial ratio distributions of Es and BCGs would have been even larger.}

This result is strongly at variance with that obtained by
RLP93, who conclude that BCGs and Es share the same axial ratio
distribution. Trying to interpret this quite evident discrepancy, we
could invoke the different procedures adopted to measure the axial
ratios. However, we should note that, among the six different values
of $q$ we have collected for our BCGs sample, the one we use to get
our distributions ($q_{_G}$) is by far the one producing the least
flattened galaxies (see Figure~\ref{axratcomp}), being also quite
consistent with $q_{_{15}}$ (the same outermost isophote used by RLP93). One
could object that, with respect to our method, the one used by RLP93
to measure $q$ tends to overweight the inner galaxy regions, which for
the BCGs are well known to be rounder than the outer ones (see again
Figure~\ref{axratcomp}). However, we should also note that our
SExtractor estimates, even if computed in a fashion similar to RLP93,
are actually in fair agreement with the isophotal axial ratios at
60~kpcs ($q_{_{60}}$; see plot 2 in Figure~\ref{axratcomp}), which
give the flattest distribution. Again, one should object that our
SExtractor estimates are computed on the whole galaxy body, while the
RLP93 ones refer to the region inside the 15~kpcs. 
In any case, it is worth stressing that, using the same axial ratio
definition ($q_{_G}$), we get very different $q$ distributions for
BCGs and normal Es. At this point, the discrepancy between our result and that
obtained by RLP93 remains rather puzzling. In Section~\ref{secdisc} we
will present a strong guess about the origin of such discrepancy. 
For now we just note that, besides the result concerning the shape
difference between BCGs and Es, the Figure~\ref{axratdistr} is telling
us something else: (i) cD galaxies are significantly flatter than
BCG\_Es (panel~b), thus implying that the BCGs do not constitute an
homogeneous class of objects; (ii) the distribution of BCG\_Es does
not significantly differ from that of normal Es (panel~c); 
(iii) the cD galaxies have an unique, characteristic shape distribution, 
regardless of the luminosity ranking in the cluster (panel~d). 
Note that this result
basically justifies why in panel~b BCG\_Es are compared
with the whole sample of cDs rather than just with BCG\_cDs.

\section{Intrinsic shapes of BCG and normal Ellipticals} \label{secrect} 

\subsection{Technicalities}

Following RLP93, hereafter we assume that the intrinsic isoluminosity
surfaces of a galaxy are similar coaxial ellipsoids with axis lengths
in the ratio 1:$\beta$:$\gamma$, with 1$\ge\beta\ge\gamma$, $\beta$
and $\gamma$ being the intrinsic axial ratios of the galaxy. Moreover,
following \citet{fran91}, we quantify the triaxiality of galaxies
through the parameter $T=(1-\beta^2)/(1-\gamma^2)$ (which becomes 0/1
for perfectly oblate/prolate bodies) and illustrate the results in the
($\beta$,$\gamma$) plane \citep[see also Figure~1 in][]{kiyi07}.

To explore the probability density funtion $\psi (\beta ,\gamma )$ of
the intrinsic axial ratios for a given sample of galaxies (hopefully
of the same morphological family), we deproject the distribution of
their observed axial ratios $\phi (q_{_G})$ using the iterative {\it
  rectification} algorithm devised by \citet{lucy74}, in particular
its equations (13), (14) and (17). In our case the equation (17) in
\citet{lucy74} becomes:
\begin{equation}
\psi^{r+1}(\beta ,\gamma ) = \frac {1}{N}\sum_{n=1}^{N}\frac {\psi^r(\beta ,\gamma )}{\phi^r(q_n)}P[q_n|\beta ,\gamma ,\epsilon (q_n)],
\end{equation}
where $N$ is the size of the galaxy sample and $r$ is the iteration
number. In this equation the $r$-th estimation of $\phi (q)$ is
obtained through the equation:
\begin{equation}
\phi^r(q_n) = \int\int \psi^r(\beta ,\gamma )P[q_n|\beta ,\gamma ,\epsilon (q_n)] d\beta d\gamma,
\end{equation}
where $P[q|\beta ,\gamma ,\epsilon (q)]$ is the probability that an
ellipsoid with intrinsic axial ratios $\beta$ and $\gamma$ is observed
with apparent axial ratio $q$, assuming a random space orientation and
the $r.m.s.$ of the errors ($\epsilon$) in the $q$ measurements to be
a function of $q$ itself.

The iterative machine requires a first guess
$\psi^0(\beta ,\gamma )$ of the probability density function we are
exploring and a choice of the error function
$\epsilon (q)$. According to \citet[][see Figure~\ref{axratcomp} and
equation~1 therein]{fasa93}, we assume the last one to be a linear
function of $q$. In particular, after some statistical comparison
among literature data, we decided to use the equation: $\epsilon
(q)=0.05(1+q)$. Concerning the choice of $\psi^0(\beta ,\gamma )$, we
tried different analytical functions, concluding the most flexible and
robust guess to be a confined normal function, bi-variate in the two
quantities $\gamma$ and $T$ (triaxiality). In this way, the first
guess of $\psi$ requires eight parameters (upper and lower limits,
central value and $\sigma$ of the normal function, for both
$T$ and $\gamma$). In general, the lower limits of
$\beta$ and $\gamma$ could be assumed to be 0.3 (strongly elongated,
prolate galaxies) and 0.05 (very thin disk galaxies),
respectively. Actually, since we are dealing with Es and cDs galaxies,
we are allowed to explore $\psi$ down to $\gamma\sim$0.3.

\begin{figure*}
\includegraphics[trim=60 0 60 0,angle=-90,scale=.65]{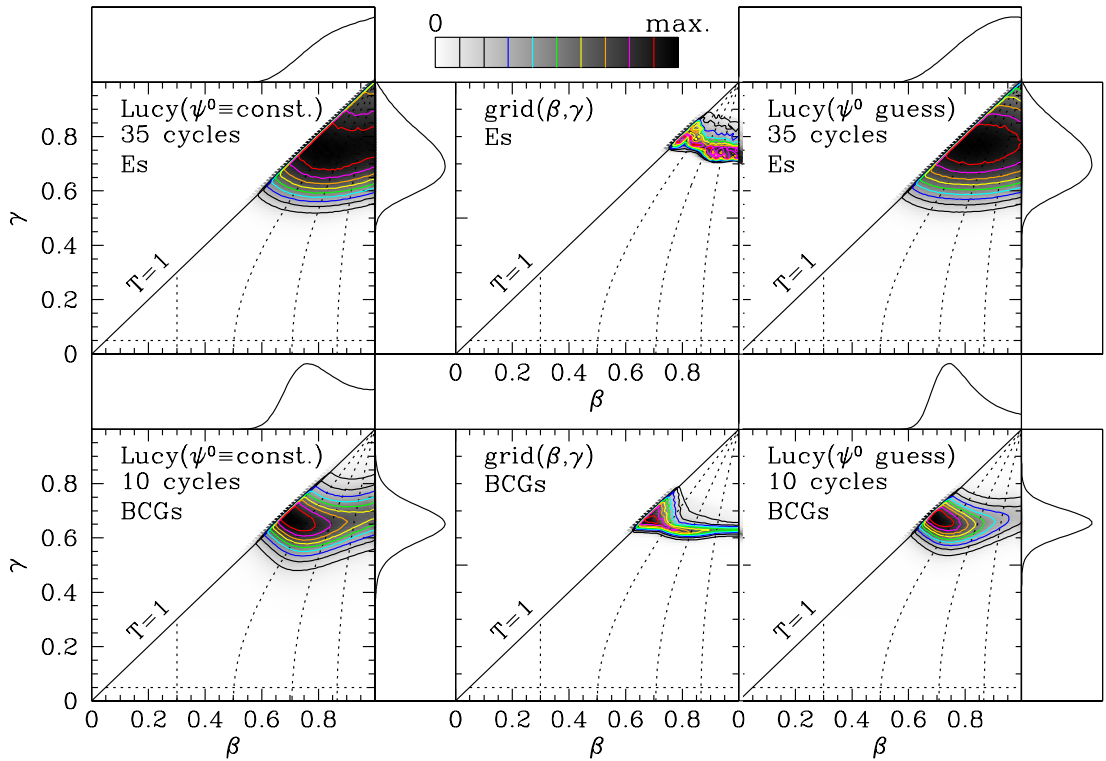}
\caption{
{\it left and right panels}: comparison between the probability
distributions $\psi (\beta ,\gamma )$ obtained with the Lucy algorithm
for the sample of normal ellipticals (top panels) and BCG
galaxies (bottom panels) in the WINGS survey. The left panels in the
figure are obtained in blind mode (initial guess
$\psi^0\equiv$const.), while in the right panels we let ourself be
guided by the results of the blind tests to guess $\psi$.
{\it middle panels}: distributions of the KS probabilities obtained,
for normal ellipticals (top panel) and BCG galaxies (bottom
panel), scanning the $(\beta ,\gamma )$ space with a bivariate,
gaussian $\psi$ and comparing in each point the expected $\phi (q)$
with the observed axis ratio distribution (see text).
The grey tones and the isophotes in each panel illustrate the density
levels (see the uppermost grey-tone scale). For the left and right
panels the marginal distributions of $\psi$ onto the $\beta$ and
$\gamma$ axes are shown in the upper and right part of each
panel, respectively. Finally, in each panel the loci of
$T$=0.25,0.5,0.75 are reported with short-dashed lines.\label{EBCGcomp}}
\end{figure*}

The left and right panels of Figure~\ref{EBCGcomp} illustrate, on the
plane ($\beta ,\gamma$), the result of the previously outlined
bi-variate Lucy's rectification procedure for the WINGS samples of Es
(top panels) and BCG galaxies (bottom panels). For each galaxy sample,
we first tried to perform the rectification without any assumption
about the initial guess ($\psi^0\equiv$const.). The distributions
obtained in this way for the two galaxy samples are shown in the left
panels of the figure. Then, we use such {\it blind} runs to refine the
results by providing more realistic initial guesses $\psi^0$,
expressed as confined, bi-variate normal functions of $\gamma$ and
$T$. The results of this refined de-projection procedure are
illustrated in the right panels of Figure~\ref{EBCGcomp}. For each sample, the
ideal number of algorithm iterations has been choosen relying upon
numerical simulations recording the cycle number for which the
provisional and the input $\psi$ distributions display the best mutual
agreement. Such ideal number turns out to depend on the sample size,
being also usually greater than the corresponding values found for the
one dimensional application of the Lucy algorithm (see Noerdlinger
1979 for perfectly oblate and prolate cases).

We are obviously aware that, since $\phi (q)$ is a function of one
variable, it cannot uniquely determine the function $\psi (\beta
,\gamma )$. This means that the previously devised Lucy algorithm
can, in principle, produce totally wrong $\psi$ distributions which
perfectly reproduce the observed distributions of $\phi (q)$. This
actually happens for certain $\psi$ distributions (i.e. disk galaxies)
when the first guess $\psi^0$ is lacking or far from the true
$\psi$. In our case, extensive numerical simulations have shown that,
even in default of the first guess ($\psi^0\equiv$const.), the
iterative Lucy machine is able to recover reasonably well the parent
$\psi$ distribution. However, in order to obtain an objective, independent
check of this conclusion, we have used a procedure similar to that adopted by
RLP93. In practice, assuming a gaussian (bi-variate) functional form of
$\psi(\beta ,\gamma )$, we have scanned the $(\beta ,\gamma )$ space
with the gaussian peak, choosing in each point the values of the two
standard deviations which, according to the KS statistic, produces the
best agreement with the observed axial ratio distribution. In this
way, we have produced a grid of KS probabilities that the $\phi (q)$
resulting from the random projection of the local gaussian $\psi$ is
drawn from the same parent population of the observed axis ratio
distribution. The middle panels of Figure~\ref{EBCGcomp} illustrate
the results of such procedure for both Es (top panel) and BCG galaxies
(bottom panel). The impressive overlap between the regions where the
distributions $\psi (\beta ,\gamma )$ peak according to the Lucy's
algorithm (left and right panels) and the regions where is highest the
probability from KS statistics (middle panels), clearly indicates
that, in our specific case, the bi-variate version of the Lucy's
algorithm works fairly well, even when it is run in blind mode
($\psi^0\equiv$const.). We also note that the distribution of KS
probabilities for the normal ellipticals in our sample (mid-top panel
in Figure~\ref{EBCGcomp}) is quite similar to the corresponding
distribution obtained by RLP93 (see Figure~5 therein).

\subsection{Results}

Figure~\ref{EBCGcomp} provides us with a robust indication about the
intrinsic shapes distributions of both Es and BCGs. Just as expected
from the observed axial ratio distributions in
Figure~\ref{axratdistr}, they turn out to be quite apart from each
other. In particular, there is a clear tendency of the BCGs towards
strong prolate configurations peaked at $\beta\sim\gamma\sim$0.67,
while the Es tend to share almost uniformely the whole range of
triaxiality, with a slight tendency toward prolateness.

\begin{figure*}
\includegraphics[trim=60 0 60 0,angle=-90,scale=.65]{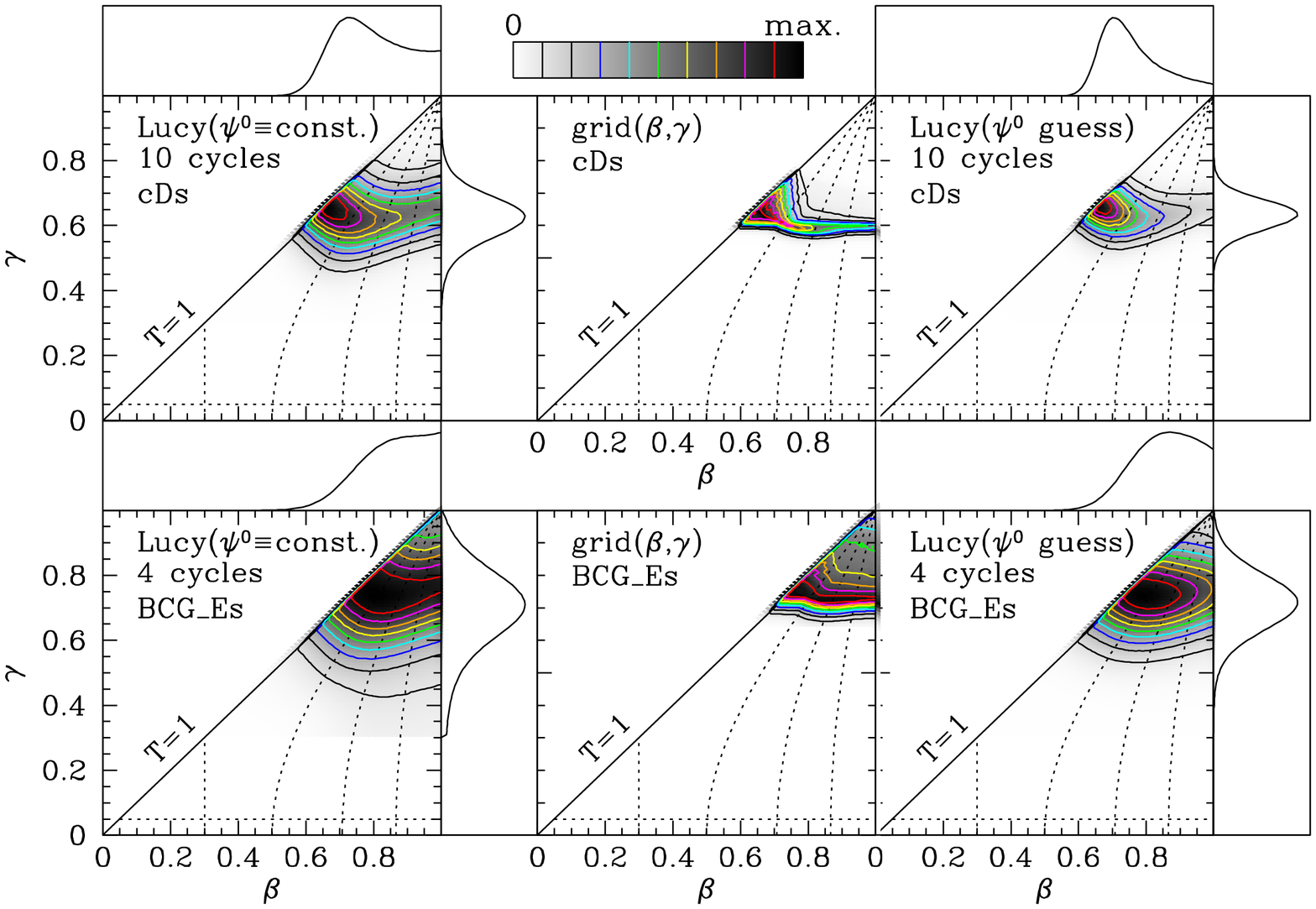}
\caption{
Comparison between the probability distributions $\psi (\beta ,\gamma )$ 
for the sample of BCG\_Es (bottom panels) and cD
galaxies (top panels) in the WINGS survey. See the caption of the
previous figure for details.\label{cDBCGcomp}}
\end{figure*}

Figure~\ref{cDBCGcomp} is similar to the previous one, but it refers
to the comparison between cD and BCG\_E galaxies in our sample. This
figure shows that the strongly prolate shape we found for the whole
population of BCGs (see lower panels of Figure~\ref{EBCGcomp}) is
entirely due to the sub-sample of cD galaxies. Moreover, in spite
of the paucity of the BCG\_E sample (responsible for the wide
distribution of the KS probabilities in the mid-bottom panel of
Figure~\ref{cDBCGcomp}), we note the similarity
between the $\psi$ distributions obtained for Es and BCG\_Es
in the top-right and bottom-right panels of
Figures~\ref{EBCGcomp} and \ref{cDBCGcomp}, respectively.

\begin{figure*}
\includegraphics[scale=.7]{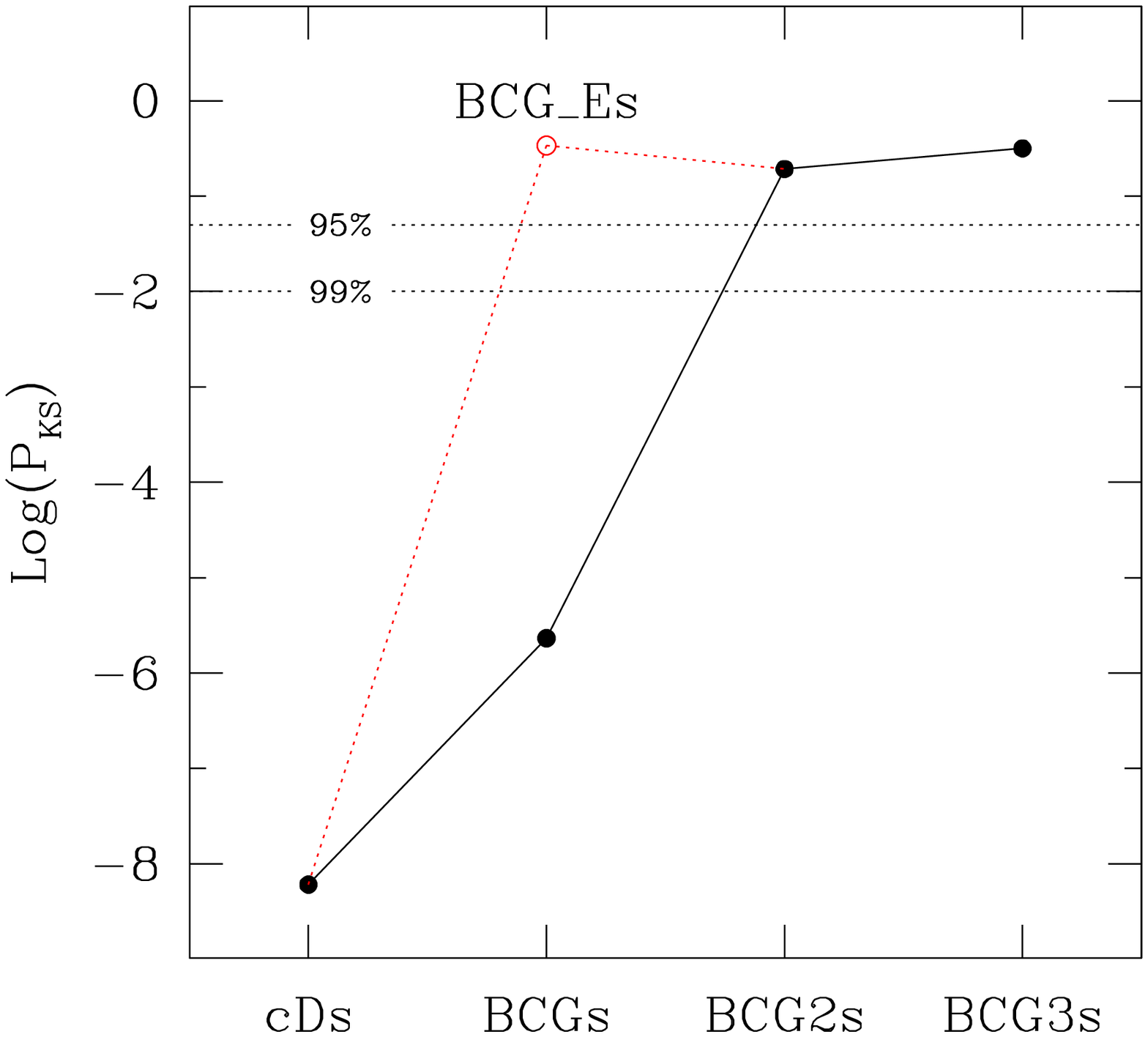}
\caption{KS probabilities relative to the comparisons of the
axial ratio distribution of Es with those of cD galaxies
and of the first three luminosity ranked ellipticals in the WINGS 
cluster. The red dot refer to the BCG\_E sample
(non cDs).\label{KSrank}}
\end{figure*}

The results illustrated in Figures~\ref{EBCGcomp} and \ref{cDBCGcomp}
suggest that cD galaxies actually constitute a foreign body inside
the global population of cluster early-type galaxies, at least as far
as the axial ratio distribution is concerned. To give more robust
statistical support to this suggestion, we have obtained the axial
ratio distributions of the second- and third-luminosity-ranked
galaxies in our cluster sample and we have compared them with the
corresponding distributions of Es, BCGs and cDs. The black dots in
Figure~\ref{KSrank} report, for the cD galaxies and for the first
three ranked ellipticals in the WINGS clusters, the KS probabilities
that the observed $\phi (q)$ of each galaxy
sample and that of the normal Es are drawn from the same parent
population.

From this figure it is evident that the $\phi (q)$ distributions of
the 2$^{nd}$- and 3$^{th}$-ranked galaxy samples (of which cDs just
represent a very small fraction) are undistinguishable from that of
Es, while for the whole sample of BCGs one can exclude the hypothesis
of common parent population between these samples and the global
elliptical sample. Nevertheless, in analogy with the $\psi$
distributions (see Figures~\ref{EBCGcomp} and
\ref{cDBCGcomp}), if we consider just the BCG\_E sample
(red, empty dot in the figure), again its $\phi (q)$
distribution turns out to be undistinguishable from that of
normal ellipticals. This confirms our previous claim about the
peculiar intrinsic shape of cDs inside the family of early-type
cluster galaxies. 

\section{Discussion} \label{secdisc}

\subsection{Cluster X-ray luminosity and BCGs shapes}

The last mentioned result could also be enough to explain the
discrepancy between our finding and the result from RLP93. In fact,
the fraction of clusters containing at least one cD galaxy in our
sample is $\sim$75\% (56 clusters). Instead, using the morphological
information provided by the {\it Nasa Extragalactic Database} (NED),
we were able to associate cD types to just $\sim$26\% of the BCGs in
the RLP93 sample. Even though most of the BCGs in the RLP93 sample are
actually lacking NED morphological information, there is an indication
that the BCG\_Es (less flattened) dominate the RLP93 sample, while the
cDs (more flattened) dominate our sample of BCGs. This circumstance
can easily justify the above discrepancy.

\begin{figure*}
\includegraphics[trim=0 150 0 150,scale=0.8]{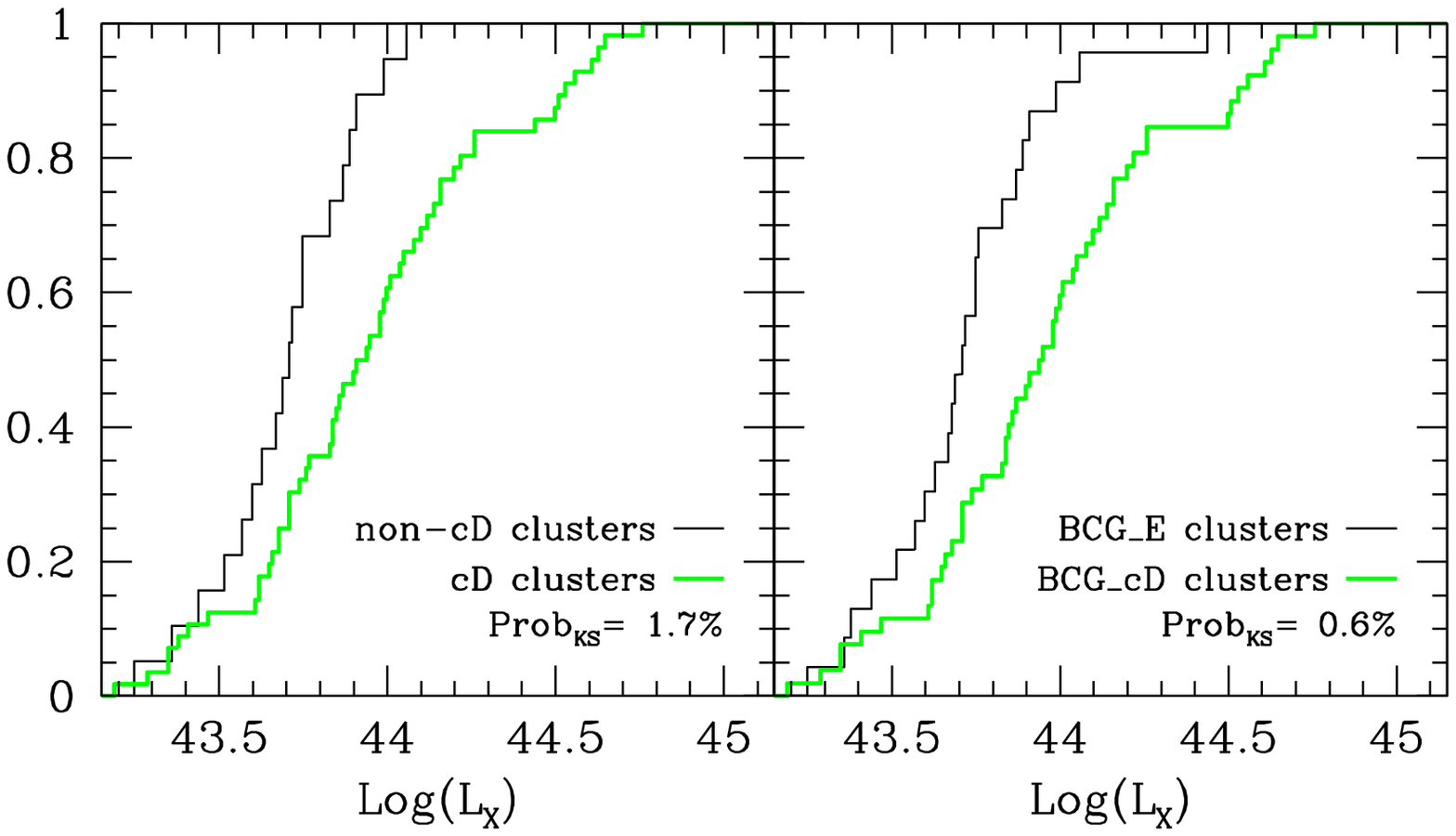}
\caption{
{\it Left panel}: comparison between the X-ray luminosity
distributions of WINGS clusters hosting (thick grey line; green in the
electronic version) and not-hosting (thin black line) at least one cD
galaxy (even if not BGC). {\it Right panel}: comparison between the
X-ray luminosity distributions of WINGS clusters with (thick grey line;
green in the electronic version) or without (thin black line)
BCG\_cDs (even if not BGC). In each panel the probability 
that the distributions
under analysis are drawn from the same parent population according to
the KS statistics is reported.
\label{lxdistr}}
\end{figure*}

Trying to explain the remarkable difference in the cD frequency
between the two cluster samples, we suggest it could be due to the
different cluster selection criteria. In fact, WINGS clusters have
been selected, in the redshift range 0.04--0.07, to be powerful X-ray emitters
(Log$L_X >$43.3~erg~s$^{-1}$). Conversely, the RLP93 sample, even spanning a
similar range of redshift, does not obey this criterion. Therefore,
the different percentages of cDs could be easily explained if cD
galaxies preferentially resided in X-ray powerful clusters. Since we
were not able to find in the literature any clear indication about
such a possibility, we tested it in our cluster sample. From
Figure~\ref{lxdistr} the tendence of cDs to preferably reside in X-ray
powerful clusters is confirmed with high significance ($>$99\%) if
just BCG\_cDs are considered (right panel), while the significance is
slightly lower if we also consider nBCG\_cDs (left panel). Such small
difference of significance might suggest that the above tendency is
stronger for BCG\_cDs than for nBCG\_cDs.  However, the KS test
applied to the X-ray luminosity distributions of clusters hosting
BCG\_cDs and nBCG\_cDs turns out to be inconclusive
(Prob$_{KS}\sim$50\%).

\begin{figure*}
\includegraphics[trim=0 150 0 150,scale=0.8]{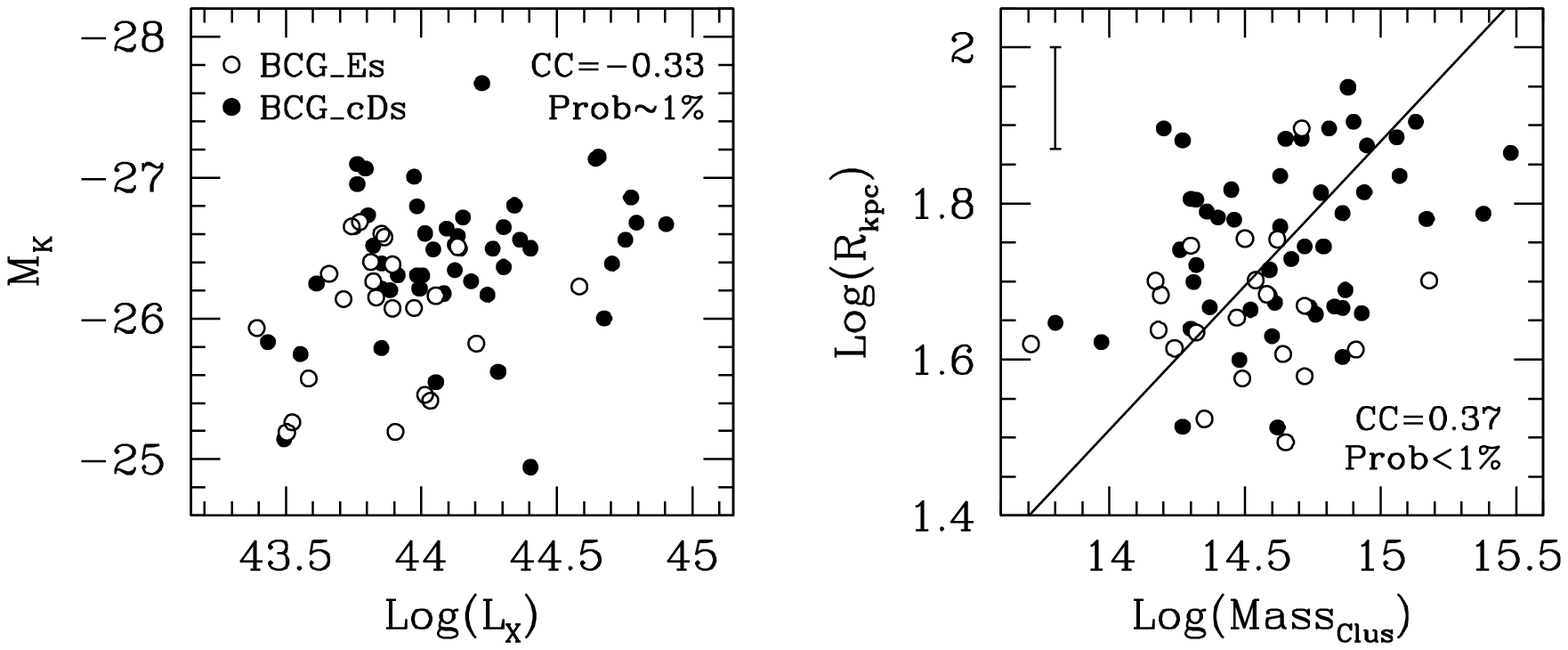}
\caption{
{\it Left panel}: correlation between absolute K-band magnitude (from
2MASS) of the BCGs in the WINGS survey and the cluster X-ray
luminosity.
{\it Right panel}: correlation between the virial cluster mass and the
equivalent radius (in kpcs) of the BCG, derived from the threshold
area. The error bar in the left-upper part of the 
plot shows the {\it r.m.s.} of the correlation (see eq.~\ref{eqmassrad} and 
text for details)
\label{massrad}}
\end{figure*}

Since cDs are known to be specially luminous and sizeable even among
the BCGs, the previous result about their preference to
reside in X-ray powerful clusters suggests that a correlation should
exist between the BCGs absolute magnitude (or size) and the cluster
X-ray luminosity (or mass). Actually, similar correlations have been
already reported in the literature
\citep{burk00,brou02,kata03,lin04,brou05,whil08}. Figure~\ref{massrad}
show the above correlations for the WINGS cluster sample. In this figure
the BCGs luminosities are expressed in K-band absolute
magnitudes from 2MASS (see Table~\ref{tab1}) and the virial masses of the
clusters are computed from the velocity dispersion of galaxies inside
them (both these quantities are reported in Table~\ref{tab1}). The
radii of the BCGs in the right panel of the figure are equivalent
radii ($\sqrt{a\times b}$) computed from the threshold area reported
in the photometric WINGS catalogs \citep{vare09}. Even if not fully 
homogeneous in the surface brightness level, this kind of size 
measurement turns out to be useful to sample the outer shapes of the 
BCGs. In Section~\ref{secdmsim} these shapes will be compared with those
of dark-matter halos from $\Lambda$CDM $N$-body simulations, which are 
hardly recoverable in their inner part, due to the insufficient resolution.
Besides the Pearson correlation coefficients, Figure~\ref{massrad}
reports the probabilities of the null hypothesis (no correlation).
Moreover, the right panel of the figure
also report the linear fit (r.m.s$\sim$0.13):

\begin{equation}
Log(R_{BCG}^{kpc})=0.37(\pm 0.056)\times Log({\rm Mass}_{Clus.})-3.67(\pm 0.81)
\label{eqmassrad}
\end{equation}

\noindent
which we use in the analysis of the $\Lambda$CDM $N$-body simulations
(see next section).

\subsection{Dark-matter halos $N$-body simulations} \label{secdmsim}

It is interesting to compare the distribution of BCG axial ratios in
the WINGS survey with that of the central part (at BCGs scale) of
cluster-sized dark-matter halos obtained with $\Lambda$CDM $N$-body
simulations. In the $\Lambda$CDM scenario, small dark matter haloes
form first and grow subsequently to larger structures via accretion
and merging processes. Such processes are generally anisotropic, so
that dark matter halos are expected to be non-spherical. A number of
papers have been devoted to the shape analysis of dark-matter halos
(see references in Section~\ref{secintro}). Recently some studies,
with both simulations and comparison with observations, have shown the
importance of the total mass in determining the final halo shape.
Indeed, it has been noted that the flattening increases with halo
mass, becoming more pronounced at $M_{vir}>10^{14}M_{\odot}$
\citep[see, among others,][]{wang08,maccio08,flores07}. Moreover, 
by using a set of hydrodynamical simulations on the cluster mass
scale, \citet{kaza04} and \citet{gott07} find that gas cooling makes
the central parts of the halos more round. Thus, one should expect
that dark matter halos of galaxy clusters are even more elongated that
the embedded central BCGs. Finally, both dynamical models
\citep{west94} and $N$-body simulations \citep{dubi98} suggest that
halos are preferentially aligned with primordial filaments, as well as
with the elongation of the host galaxy cluster.

A truly realistic comparison between our observations and numerical
simulations would require an accurate modelling of hydrodynamics,
radiative cooling, star formation and energy feedback from SNs and
AGNs, high enough mass and force resolution, and full control over the
numerical effects. Unfortunately, our incomplete knowledge of the
details of galaxy formation, and the insufficient computational
capabilities, make this impossible as of today. We must thus rely on
the more robust - although simpler - modelling of the dark-matter
distribution. In order to make such a comparison still meaningful, we
chose to compare the intrinsic shape of BCGs with that
of the innermost region of cluster-sized dark matter halos extracted
from a cosmological simulation.

We compare the intrinsic shape of our BCGs with that of the innermost
region of cluster-sized dark matter haloes extracted from GIF2
simulation \citep[see][, hereafter GIF2]{gao04}
\footnote{data publicly available at
$http://www.mpa-garching.mpg.de/Virgo/data\_download.html$}, 
a cosmological $N$-body simulation of the concordance $\Lambda$CDM
model, performed with $400^3$ dark-matter particles in a box of $110$
Mpc/h on a side. Halos were identified at redshift $z=0$ using a
spherical overdensity criterion, cut at the $\Lambda$CDM virial
overdensity: $\Delta_{vir} = 324$ \citep{eke96}. Among these we
selected 3510 halos with virial mass above $10^{14}M_\odot$,
appropriate to compare to the WINGS galaxy clusters (see lower mass  
limit in right panel of Figure~\ref{massrad}). 

\begin{figure*}
\includegraphics[scale=0.8]{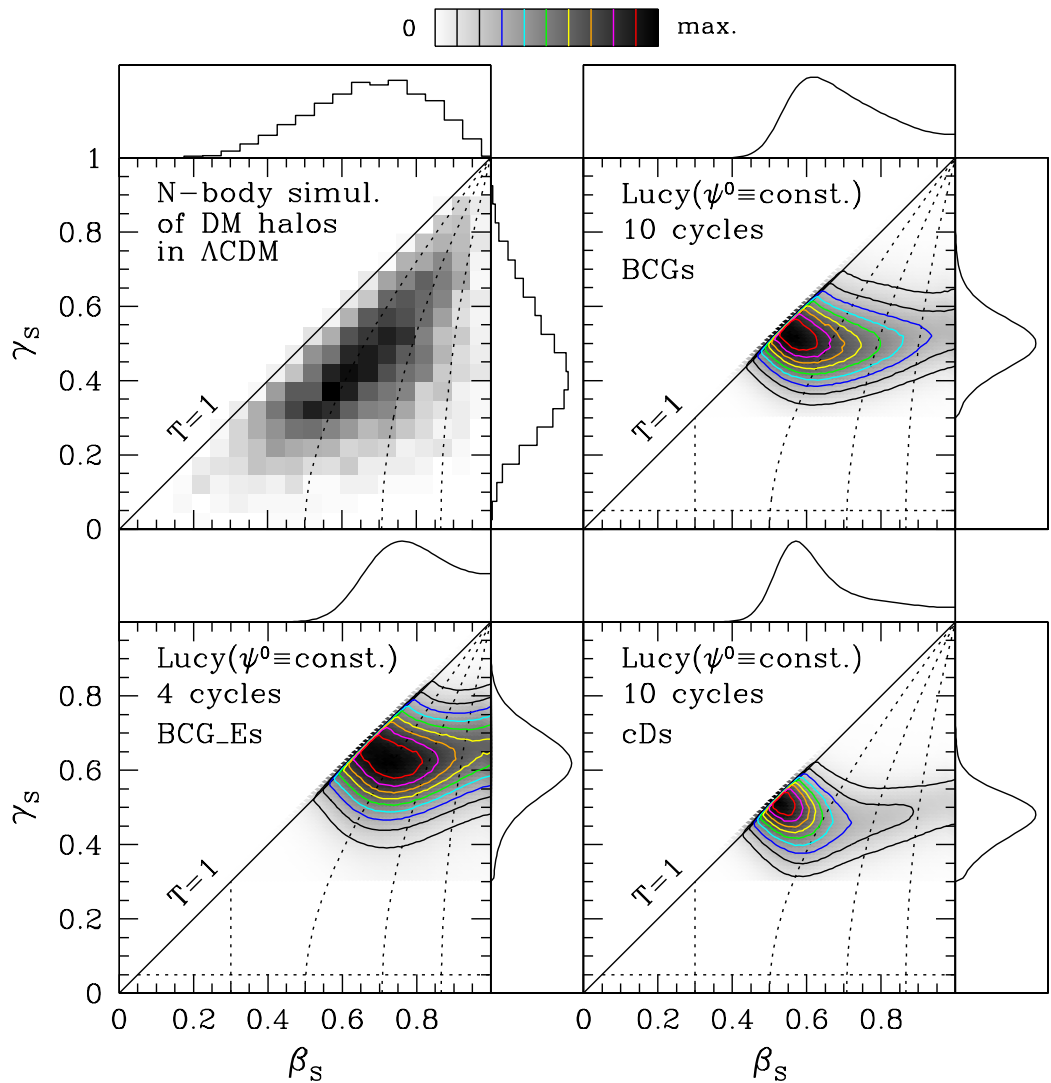}
\caption{
Distribution of the intrinsic axial ratios for the sample
of dark-matter halos extracted from the GIF2 $N$-body simulation
(upper-left panel), compared with the corresponding distributions
($\beta_{_S},\gamma_{_S}$)
obtained deprojecting the distributions of the threshold isophotes 
$q_{_S}$ for our samples of BCGs (upper-right), BCG\_Es (bottom-left) and
cDs (bottom-right).
\label{nbodydistr}}
\end{figure*}

Given the discrete structure (particles) of the matter distribution in
the $N$-body simulations, we have used an iterative procedure to
obtain the shapes of the halo isodensity shells. As a first guess, we
calculated the main axes of the dark-matter particle distribution
inside a sphere of radius $R_{BCG}$, where $R_{BCG}$ is randomly
assigned from equation~(3), assuming a gaussian scatter (see also
right panel of Figure~\ref{massrad}). Then, keeping the volume of the
region fixed, we define a new ellipsoidal region using these axes and
obtain new values for the main axes. We iterate this procedure until
the main axes converge to a stable value, which we assigned to the
dark matter halo as representative of its ``simulated'' BCG. The
typical dark matter overdensity at the radii considered in this
procedure is of order $10^5$ times the background density. In
the upper-left panel of
Figure~\ref{nbodydistr} we plot the distribution of the intrinsic
axial ratios for the dark matter halo regions thus identified.

The comparison of this plot with Figures~\ref{EBCGcomp} and \ref{cDBCGcomp},
led us to the conclusion that dark matter halos are even more
elongated and strongly prolate than our BCGs/cDs galaxies. However, it
is worth recalling that the axial ratios in Figures~\ref{EBCGcomp} and
\ref{cDBCGcomp} refer to the effective isophotes, while the size
distribution we use to extract the intrinsic shapes of dark-matter
halos in the GIF2 simulation has been calibrated (right panel of
Figure~\ref{massrad} and eq.~\ref{eqmassrad}) on the threshold
isophotes of the BCG sample, which, as seen in panel 15 of
Figure~\ref{axratcomp}, tend to be flatter than the effective
isophotes. To obtain a more consistent comparison, in the remaining 
panels of Figure~\ref{nbodydistr} we illustrate the
results of the previously outlined rectification procedure (with
initial guess $\psi^0\equiv$const.) on our samples of BCGs
(upper-right), BCG\_Es (bottom-left) and cDs (bottom-right), when the
axial ratios of the threshold isophotes ( $q_{_S}$) are used instead
of the effective axial ratios ( $q_{_G}$). Even if in this case the
BCGs tend to be more elongated than in the case of the effective
isophote, still they turn out to be rounder when compared to the dark
matter halos, giving support to the results of \citet{kaza04} and
\citet{gott07} about gas cooling effects on the central parts of
simulated galaxy clusters.

\subsection{The shapes of BCGs and their evolution}

In our sample of BCGs, the distributions 
of many quantities indicate that BCG\_cDs are systematically
different from BCG\_Es (stronger flattening, higher mass and larger size, 
preference to reside in massive and X-ray powerful clusters;
see Figures~\ref{axratdistr}, \ref{lxdistr} and \ref{massrad}).
However, the only statistically robust differences concern the shapes
($P_{KS}<$0.001) and the X-ray luminosity of the host clusters 
($P_{KS}<$0.01).

The tendency of cDs to reside in more X-ray powerful clusters and
their strong prolateness when compared to BCG\_Es, together with the
(even more) elongated and prolate shapes of the inner parts of
cluster-sized, simulated dark matter halos (see
Figure~\ref{nbodydistr}), could give interesting hints for
understanding the formation histories of both the BCGs and the host
clusters themselves.

A detailed discussion about what the above mentioned results, together
with those emerging from recent dynamical and stellar population
studies of BCGs, imply about their formation and evolution, is beyond
the scope of this paper. We actually plan going deeper into this
topic in a forthcoming paper, using the whole information on BCGs from
the WINGS database. We just provide here a few speculations, also
based on the results of some recent analyses in the literature.

\citet{stot08}, \citet{whil08} and \citet{coll09} have demonstrated 
that the stellar population in BCGs has been mostly in place (90\% of
the total mass) since z$\sim$2 and that, contrary to the predictions
of cosmological simulations and hierarchical-based galaxy formation
models \citep{bowe06,delu07}, the BCGs are almost fully assembled a
few billion years after the Big Bang. On the other hand, converging
indications exist that the most luminous (and massive) BCGs may often
display signatures of most recent star formation (blue-cores), likely
related to the presence in the host clusters of cooling flows and
X-ray luminosity excess with respect to the average $L_X-T_X$ relation
\citep{bild08,loub09}. The most luminous BCGs are also claimed by
\citet{bild08} to be closer to the peaks of the cluster X-ray emission
with respect to the less luminous BCGs. Finally, \citet{cozi09} show
that most BCGs are not at rest in the potential well of their
clusters, suggesting a merging-group scenario, where BCGs formed first
in smaller subsystems and clusters formed more recently from the
mergers of many such groups. They also claim that the
relative peculiar velocity inside the cluster (i.e. the difference
between specific and average velocity, normalized to the velocity
dispersion) is lower for BCG\_cDs than for BCG\_Es.

All these claims, together with the findings reported in the present
paper and the fact that the BCG\_cDs are (on average) more luminous
and massive than the BCG\_Es, point toward a scenario in which
BCG\_cDs and BCG\_Es have experienced quite different formation and
evolution processes, although both have been assembled very early
(old stellar populations). In particular, the BCG\_cDs could have been
formed close to the centers of quite sizeable and prolate
\citep{gott07} dark-matter halos, dominant with respect to the
surroundig ones and progenitors of the most massive, present day halos
(small displacement of the BCGs from the cluster centers and small
relative peculiar velocity). In this context, considerable and
prolonged accretion of material through cooling flows (enhanced X-ray
emission) and/or dry minor mergers \citep{bern09}, associated with
modest star formation (blue-cores), could have been at work from
preferential directions (the streaming filaments), thus producing a
puffing up process along them, leading to the formation of extended and
elongated halos (cD morphology). On the other side, the BCG\_Es could
have formed inside group-sized halos, not much larger than the surrounding 
ones. Later on, according to the scenario proposed by \citet{cozi09}, the
galaxy groups originating the BCG\_Es could have merged with many
other groups, to form cluster-sized systems. In this context, one
should expect the BCGs to show very low or absent star formation, to have larger
displacement from the cluster centers and larger relative peculiar
velocities compared to the BCG\_cDs. Moreover, in this case
the host clusters are expected to show less frequently cooling flows
and to be less powerful X-ray emitters.

\section{Summary} \label{secsumm}

In this paper we have analysed the apparent axial ratio distributions
of the BCGs and of the normal ellipticals in our sample of 75
galaxy clusters from the WINGS survey. Most BCGs in our clusters (52)
have been classified as cD galaxies. The cD sample has been completed
by 14 additional cDs (non-BCGs) we found in our clusters.
We have deprojected the apparent axial ratios distributions of BCGs, cDs and
normal ellipticals using a bi-variate version of the rectification Lucy's
algorithm. Since in our case the bi-variate distribution of the intrinsic
axial ratios cannot be uniquely determined by the univariate distribution
of the apparent axial ratios, we have used an independent Monte-Carlo
technique to support the results of the Lucy's algorithm. Finally, we have
used the GIF2 $\Lambda$CDM $N$-body simulations of cluster-sized, dark-matter
halos, to compare the intrinsic shapes of the inner part (BCG-sized) of halos 
with those of the observed BCGs.

The main conclusions are the following:

\begin{itemize}

\item 
The {\it normal} Es have triaxial shape, the
triaxiality parameter sharing almost evenly the ($\beta$,$\gamma$)
space (with a slight preference for prolateness) and the $\gamma$
parameter peaking around 0.7. In this case, our result is fairly in
agreement with that provided by RLP93;

\item
the normal BCGs (non-cDs) and the second and third luminosity
ranked Es in the clusters do not differ from the global
population of Es as far as the distribution of the intrinsic 
shapes is concerned;

\item
the cDs have triaxial shape too. However, in this case the tendency
towards prolateness is very strong and the preferred values of $\beta$
and $\gamma$ are significantly lower ($\sim$0.6) than the peak values found
for Es;

\item
since more than 2/3 of the BCGs in our sample are cD galaxies, the
results of the shape analysis for our global sample of BCGs are
similar to those obtained for cDs (prolateness, significant difference 
from the Es);

\item
this result turns out to be strongly at variance with the conclusions
given by RLP93, who found that BCGs and Es have similar apparent and
intrinsic axial ratio distributions. This discrepancy is harldy
attributable just to systematic differences in the axial ratios
measurements;

\item
since we find that, among BCGs, cDs are quite flatter that non-cDs,
and since the WINGS clusters have been selected to be X-ray powerful,
while the RLP93 clusters haven't, we suggest the above discrepancy to
be caused by a preference of cDs to reside in X-ray emitting
clusters. Actually, this hypothesis turns out to be supported
by the comparison between the X-ray luminosity distributions of the cD
and non-cD clusters in our sample;

\item
the prolateness of the BCGs (in particular of the cDs) could reflect the 
shape of the associated dark-matter halos, according to the GIF2 $N$-body data.

\end{itemize}

\section*{Acknowledgments}

The simulations in this paper were carried out by the Virgo
Supercomputing Consortium using computers based at Computing Centre of
the Max-Planck Society in Garching and at insitute for Computational
Cosmology. The data are publicly available at
www.mpa-garching.mpg.de/NumCos.

We thank Paola Mazzei and Marisa Girardi for useful discussions. We are 
also grateful to the anonymous referee, whose comments allowed us to 
improve the paper.

\begin{landscape}
\begin{table}
\begin{minipage}{250mm}
\caption{Main properties of the BCG sample and of the host clusters}
\label{tab1}
\begin{tabular}{cccccccccccccccc}
\hline\hline
Cluster & Log$(L_X)^{(a)}$ & Log$(\Sigma)^{(b)}$ & 
z$^{(c)}$ & WINGS\_ID$^{(d)}$ &Ty$^{(e)}$ & 
$M_V^{(f)}$ & $M_K^{(g)}$ & R$_e^{(h)}$ & 
$n^{(i)}$ & $q_{_G}^{(j)}$ & $q_{_{15}}^{(k)}$ & 
$q_{_{30}}^{(l)}$ & $q_{_{60}}^{(m)}$ & $q_{_S}^{(n)}$ & 
$q_{_L}^{(o)}$ \\
 & 10$^{44}$ergs~s$^{-1}$ & km~s$^{-1}$ & & & & & & kpc & & & & & & & \\
\hline
      A85 &  4.28 &  3.022 &   0.0551 & WINGSJ004150.45-091811.5 & cD &  -23.89 &  -26.89 &  33.86 &   2.5 &   0.707 &   0.770 &   0.710 &   0.640 &   0.610 &   0.641  \\
     A119 &  1.65 &  2.936 &   0.0444 & WINGSJ005616.12-011519.0 & cD &  -23.45 &  -26.58 &  25.67 &   3.6 &   0.769 &   0.740 &   0.690 &   0.520 &   0.545 &   0.767 \\
     A133 &  1.82 &  2.908 &   0.0566 & WINGSJ010241.72-215255.4 & cD &  -23.48 &  -26.52 &  28.75 &   3.0 &   0.666 &   0.680 &   0.590 &   0.550 &   0.498 &   0.527 \\
     A147 &  0.28 &  2.823 &   0.0447 & WINGSJ010812.04+021138.2 &  E &  -22.96 &  -25.61 &  26.44 &   4.4 &   0.769 &   0.770 &   0.770 &   0.770 &   0.789 &   0.776 \\
     A151 &  0.52 &  2.881 &   0.0532 & WINGSJ010851.13-152423.0 &  E &  -24.08 &  -26.58 &  52.03 &   5.0 &   0.773 &   0.800 &   0.720 &   0.760 &   0.789 &   0.706 \\
     A160 &  0.19 &  2.749 &   0.0438 & WINGSJ011259.57+152928.8 & cD &  -22.99 &  -25.89 &  26.78 &   3.7 &   0.780 &   0.800 &   0.720 &   0.720 &   0.691 &   0.716 \\
     A168 &  0.56 &  2.702 &   0.0450 & WINGSJ011457.58+002551.1 &  E &  -23.16 &  -26.06 &  21.47 &   4.8 &   0.799 &   0.800 &   0.700 &   0.670 &   0.620 &   0.708 \\
     A193 &  0.79 &  2.880 &   0.0485 & WINGSJ012507.64+084157.2 & cD &  -24.67 &  -26.52 &        &       &   0.879 &   0.880 &   0.730 &   0.760 &   0.752 &   0.589 \\
     A311 &  0.41 &        &   0.0661 & WINGSJ020928.41+194636.2 & cD &  -23.59 &  -26.94 &  43.09 &   4.7 &   0.576 &   0.630 &   0.520 &   0.460 &   0.445 &   0.519 \\
     A376 &  0.71 &  2.930 &   0.0476 & WINGSJ024603.94+365419.1 & cD &  -23.55 &  -26.25 &  49.09 &   5.9 &   0.876 &   0.890 &   0.790 &   0.700 &   0.693 &   0.869 \\
     A500 &  0.72 &  2.818 &   0.0678 & WINGSJ043852.51-220639.0 & cD &  -23.55 &  -26.26 &  44.62 &   7.0 &   0.780 &   0.760 &   0.620 &   0.630 &   0.662 &   0.664 \\
    A548b &  0.15 &  2.928 &   0.0416 & WINGSJ054529.62-255556.8 & cD &  -21.57 &         &   4.88 &   3.2 &   1.021 &   0.940 &   0.790 &   0.790 &   0.847 &   0.460 \\
     A602 &  0.57 &  2.857 &   0.0619 & WINGSJ075326.61+292134.4 &  E &  -22.49 &  -25.14 &  23.37 &   3.0 &   0.832 &   0.800 &   0.760 &   0.760 &   0.721 &   0.589 \\
     A671 &  0.45 &  2.957 &   0.0507 & WINGSJ082831.66+302553.0 & cD &  -24.04 &  -26.72 &  51.55 &   5.5 &   0.766 &   0.790 &   0.720 &   0.670 &   0.675 &   0.698 \\
     A754 &  4.09 &  3.000 &   0.0547 & WINGSJ090832.39-093747.3 & cD &  -23.99 &  -26.58 &  47.72 &   4.5 &   0.721 &   0.700 &   0.730 &   0.650 &   0.651 &   0.776 \\
     A780 &  3.38 &  2.866 &   0.0539 & WINGSJ091805.68-120543.2 & cD &  -23.66 &  -26.04 &  39.18 &   4.4 &   0.838 &   0.820 &   0.800 &   0.730 &   0.742 &   0.537 \\
     A957 &  0.40 &  2.851 &   0.0451 & WINGSJ101338.27-005531.2 &  E &  -23.69 &  -26.63 &  28.91 &   4.3 &   0.877 &   0.810 &   0.790 &   0.790 &   0.777 &   0.710 \\
     A970 &  0.77 &  2.883 &   0.0591 & WINGSJ101725.71-104120.2 &  E &  -23.18 &  -25.40 &        &       &   0.856 &   0.720 &   0.750 &   0.780 &   0.788 &   0.589 \\
    A1069 &  0.48 &  2.839 &   0.0653 & WINGSJ103943.44-084112.3 &  E &  -23.62 &  -26.12 &  34.87 &   4.5 &   0.868 &   0.760 &   0.870 &   0.870 &   0.875 &   0.735 \\
    A1291 &  0.22 &  2.632 &   0.0509 & WINGSJ113223.22+555803.0 & cD &  -22.79 &  -25.17 &  37.33 &   4.2 &   0.745 &   0.840 &   0.570 &   0.580 &   0.513 &   0.671 \\
   A1631a &  0.37 &  2.806 &   0.0461 & WINGSJ125318.41-153203.8 &  E &  -23.35 &  -26.20 &  31.15 &   5.8 &   0.728 &   0.740 &   0.700 &   0.710 &   0.689 &   0.708 \\
    A1644 &  0.04 &  3.033 &   0.0467 & WINGSJ125711.60-172434.0 & cD &  -23.94 &  -25.14 &  44.93 &   2.9 &   0.638 &   0.630 &   0.590 &   0.520 &   0.525 &   0.557 \\
    A1668 &  0.81 &  2.812 &   0.0634 & WINGSJ130346.60+191617.4 &  E &  -23.27 &  -26.17 &  26.58 &   3.2 &   0.751 &   0.760 &   0.730 &   0.730 &   0.714 &   0.895 \\
    A1736 &  1.21 &  2.931 &   0.0458 & WINGSJ132644.09-272621.8 & cD &  -23.41 &  -27.65 &        &       &   0.747   &   0.670 &   0.630 &   0.630 &   0.620 &   0.513 \\
    A1795 &  5.67 &  2.860 &   0.0633 & WINGSJ134852.51+263534.5 & cD &  -23.89 &  -26.66 &  52.00 &   4.2 &   0.707 &   0.760 &   0.740 &   0.540 &   0.531 &   0.603 \\
    A1831 &  0.97 &  2.735 &   0.0634 & WINGSJ135915.11+275834.5 & cD &  -24.49 &  -27.00 &  95.06 &   5.9 &   0.689 &   0.730 &   0.650 &   0.540 &   0.572 &   0.530 \\
    A1983 &  0.24 &  2.722 &   0.0447 & WINGSJ145255.33+164210.5 &  E &  -22.64 &  -25.23 &  27.83 &   4.5 &   0.934 &   0.920 &   0.910 &   0.880 &   0.856 &   0.955 \\
    A1991 &  0.69 &  2.777 &   0.0584 & WINGSJ145431.50+183832.8 & cD &  -23.32 &  -26.20 &  30.85 &   3.1 &   0.658 &   0.700 &   0.640 &   0.590 &   0.571 &   0.641 \\
    A2107 &  0.56 &  2.772 &   0.0410 & WINGSJ153938.92+214658.1 &  E &  -23.41 &  -26.43 &  24.95 &   3.1 &   0.787 &   0.810 &   0.710 &   0.710 &   0.715 &   0.295 \\
    A2124 &  0.69 &  2.904 &   0.0666 & WINGSJ154459.02+360633.9 & cD &  -23.77 &  -26.77 &  36.33 &   3.2 &   0.737 &   0.780 &   0.720 &   0.650 &   0.629 &   0.659 \\
    A2149 &  0.42 &  2.548 &   0.0679 & WINGSJ160128.11+535650.3 &  E &  -24.29 &  -26.60 &        &       &   0.788 &   0.850 &   0.780 &   0.760 &   0.711 &   0.791 \\
    A2169 &  0.23 &  2.707 &   0.0578 & WINGSJ161358.09+491122.3 &  E &  -23.25 &  -25.28 &        &       &   0.741 &   0.720 &   0.720 &   0.660 &   0.629 &   0.710 \\
    A2256 &  3.60 &  3.105 &   0.0581 & WINGSJ170427.22+783825.4 & cD &  -23.60 &  -26.44 &  27.43 &   3.5 &   0.866 &   0.840 &   0.830 &   0.850 &   0.898 &   0.883 \\
    A2271 &  0.32 &  2.702 &   0.0576 & WINGSJ171816.66+780106.2 &  E &  -23.75 &  -26.25 &  56.14 &   4.8 &   0.709 &   0.760 &   0.670 &   0.670 &   0.634 &   0.676 \\
    A2382 &  0.46 &  2.948 &   0.0641 & WINGSJ215155.62-154221.2 &  E &  -23.29 &  -26.47 &  35.33 &   5.1 &   1.058 &   0.970 &   0.920 &   0.920 &   0.828 &   0.513 \\
    A2399 &  0.51 &  2.852 &   0.0578 & WINGSJ215701.72-075022.0 & cD &  -22.84 &  -25.96 &  13.19 &   3.2 &   0.719 &   0.740 &   0.780 &   0.780 &   0.845 &   0.746 \\
    A2415 &  0.86 &  2.843 &   0.0575 & WINGSJ220526.12-054431.1 & cD &  -22.90 &  -26.14 &  20.75 &   5.2 &   0.731 &   0.720 &   0.620 &   0.620 &   0.594 &   0.592 \\
    A2457 &  0.73 &  2.763 &   0.0584 & WINGSJ223540.81+012905.8 & cD &  -23.53 &  -26.59 &  36.83 &   5.2 &   0.684 &   0.740 &   0.590 &   0.580 &   0.583 &   0.662 \\
   A2572a &  0.52 &  2.800 &   0.0390 & WINGSJ231711.95+184204.7 &  E &  -21.31 &  -26.64 &   3.12 &   2.5 &   0.901 &   0.880 &   0.510 &   0.570 &   0.577 &   0.849 \\
    A2589 &  0.95 &  2.912 &   0.0419 & WINGSJ232357.44+164638.3 & cD &  -23.83 &  -26.30 &  61.08 &   5.1 &   0.707 &   0.740 &   0.570 &   0.410 &   0.408 &   0.501 \\
    A2593 &  0.59 &  2.846 &   0.0417 & WINGSJ232420.08+143849.8 & cD &  -23.05 &  -26.30 &  21.29 &   2.8 &   0.650 &   0.660 &   0.620 &   0.610 &   0.613 &   0.635 \\
    A2622 &  0.55 &  2.843 &   0.0610 & WINGSJ233501.47+272220.9 & cD &  -23.14 &  -26.22 &  24.05 &   4.2 &   0.703 &   0.720 &   0.650 &   0.610 &   0.582 &   0.646 \\
    A2626 &  0.99 &  2.796 &   0.0548 & WINGSJ233630.49+210847.3 & cD &  -23.27 &  -26.48 &  25.62 &   2.3 &   0.666 &   0.720 &   0.670 &   0.630 &   0.618 &   0.723 \\
    A2657 &  0.82 &  2.581 &   0.0402 & WINGSJ234457.42+091135.2 & cD &  -22.66 &  -25.56 &  32.60 &   3.5 &   0.691 &   0.650 &   0.630 &   0.640 &   0.628 &   0.631 \\
 \hline

\end{tabular}
\end{minipage}
\end{table}
\end{landscape}

\setcounter{table}{0}
\begin{landscape}
\begin{table}
\begin{minipage}{250mm}
\caption{(continue) Main properties of the BCG sample and of the host clusters}
\label{tab1}
\begin{tabular}{cccccccccccccccc}
\hline\hline
Cluster & Log$(L_X)^{(a)}$ & Log$(\Sigma)^{(b)}$ & 
z$^{(c)}$ & WINGS\_ID$^{(d)}$ &Ty$^{(e)}$ & 
$M_V^{(f)}$ & $M_K^{(g)}$ & R$_e^{(h)}$ & 
$n^{(i)}$ & $q_{_G}^{(j)}$ & $q_{_{15}}^{(k)}$ & 
$q_{_{30}}^{(l)}$ & $q_{_{60}}^{(m)}$ & $q_{_S}^{(n)}$ & 
$q_{_L}^{(o)}$ \\
 & 10$^{44}$ergs~s$^{-1}$ & km~s$^{-1}$ & & & & & & kpc & & & & & & & \\
\hline
   A2665 &  0.97 &        &   0.0556 & WINGSJ235050.55+060858.9 &  E &  -23.59 &  -26.53 &  34.02 &   3.4 &   0.797 &   0.830 &   0.780 &   0.720 &   0.742 &   0.000 \\
    A2717 &  0.52 &  2.743 &   0.0490 & WINGSJ000312.95-355613.3 & cD &  -23.56 &  -26.24 &  42.48 &   4.7 &   0.921 &   0.870 &   0.940 &   0.930 &   0.932 &   0.979 \\
    A2734 &  1.30 &  2.744 &   0.0625 & WINGSJ001121.64-285115.5 & cD &  -23.44 &  -26.47 &  24.66 &   5.2 &   0.856 &   0.760 &   0.610 &   0.630 &   0.513 &   0.621 \\
    A3128 &  2.71 &  2.946 &   0.0600 & WINGSJ032950.60-523446.8 & cD &  -24.18 &  -26.42 &        &       &   0.928 &   0.880 &   0.810 &   0.780 &   0.697 &   0.731 \\
    A3158 &  2.71 &  3.036 &   0.0593 & WINGSJ034329.69-534131.7 &  E &  -23.41 &  -26.34 &  22.37 &   4.1 &   0.922 &   0.920 &   0.900 &   0.920 &   0.888 &   0.851 \\
    A3266 &  3.14 &  3.136 &   0.0593 & WINGSJ043113.27-612711.9 & cD &  -24.51 &  -27.20 &  98.44 &   6.2 &   0.649 &   0.700 &   0.550 &   0.540 &   0.548 &   0.548 \\
    A3376 &  1.27 &  2.892 &   0.0461 & WINGSJ060041.09-400240.4 & cD &  -23.32 &  -26.20 &  28.11 &   4.7 &   0.669 &   0.670 &   0.630 &   0.620 &   0.599 &   0.635 \\
    A3395 &  1.43 &  2.898 &   0.0500 & WINGSJ062736.25-542657.9 & cD &  -23.29 &  -26.31 &  46.78 &   4.1 &   0.478 &   0.470 &   0.410 &   0.420 &   0.447 &   0.381 \\
    A3490 &  0.88 &  2.841 &   0.0688 & WINGSJ114520.15-342559.3 & cD &  -23.72 &  -26.61 &  62.34 &   6.8 &   0.661 &   0.590 &   0.570 &   0.590 &   0.540 &   0.603 \\
    A3497 &  0.74 &  2.861 &   0.0680 & WINGSJ115946.30-313141.6 &  E &  -22.45 &  -25.47 &  13.19 &   5.5 &   0.857 &   0.710 &   0.730 &   0.730 &   0.737 &   0.755 \\
   A3528a &  0.68 &  2.954 &   0.0535 & WINGSJ125441.01-291339.5 & cD &  -23.77 &  -27.15 &  13.08 &   2.2 &   0.956 &   0.990 &   0.940 &   0.940 &   0.910 &   0.834 \\
   A3528b &  1.01 &  2.936 &   0.0535 & WINGSJ125422.23-290046.8 & cD &  -24.04 &  -26.74 &  69.87 &   6.7 &   0.695 &   0.690 &   0.620 &   0.620 &   0.572 &   0.634 \\
    A3530 &  0.44 &  2.751 &   0.0537 & WINGSJ125535.99-302051.3 & cD &  -24.17 &  -27.07 &  85.40 &   5.7 &   0.713 &   0.500 &   0.560 &   0.620 &   0.659 &   0.488 \\
    A3532 &  1.44 &  2.793 &   0.0554 & WINGSJ125721.97-302149.1 & cD &  -24.55 &  -26.60 &        &       &   0.777 &   0.770 &   0.690 &   0.720 &   0.769 &   0.000 \\
    A3556 &  0.48 &  2.747 &   0.0479 & WINGSJ132406.71-314011.6 & cD &  -23.53 &  -26.52 &  27.62 &   5.2 &   0.689 &   0.670 &   0.790 &   0.820 &   0.804 &   0.594 \\
    A3558 &  3.20 &  2.961 &   0.0480 & WINGSJ132756.84-312943.9 & cD &  -23.97 &  -27.10 &  34.63 &   2.8 &   0.727 &   0.720 &   0.670 &   0.460 &   0.463 &   0.618 \\
    A3560 &  0.67 &  2.851 &   0.0489 & WINGSJ133225.76-330808.9 &  E &  -21.99 &  -26.07 &        &       &   0.781 &         &         &         &   0.680 &   0.000 \\
    A3667 &  4.47 &  2.997 &   0.0556 & WINGSJ201227.32-564936.3 & cD &  -23.98 &  -26.69 &  41.44 &   4.0 &   0.914 &   0.950 &   0.790 &   0.620 &   0.572 &   0.689 \\
    A3716 &  0.52 &  2.921 &   0.0462 & WINGSJ205156.94-523746.8 & cD &  -23.77 &  -26.80 &  32.60 &   4.2 &   0.739 &   0.740 &   0.660 &   0.640 &   0.642 &   0.525 \\
    A3809 &  1.15 &  2.751 &   0.0627 & WINGSJ214659.07-435356.2 &  E &  -23.11 &  -25.82 &  23.13 &   3.2 &   0.782 &   0.830 &   0.710 &   0.730 &   0.720 &   0.783 \\
    A3880 &  0.95 &  2.883 &   0.0584 & WINGSJ222754.43-303431.8 & cD &  -23.18 &  -26.51 &  20.22 &   2.6 &   0.889 &   0.880 &   0.740 &   0.710 &   0.609 &   0.793 \\
    A4059 &  1.58 &  2.854 &   0.0475 & WINGSJ235700.71-344532.8 & cD &  -23.95 &  -26.87 &  38.99 &   3.0 &   0.676 &   0.680 &   0.620 &   0.580 &   0.506 &   0.600 \\
  IIZW108 &  1.12 &  2.710 &   0.0483 & WINGSJ211355.90+023355.4 & cD &  -23.78 &         &  46.30 &   2.8 &   0.627 &   0.690 &   0.620 &   0.510 &   0.527 &   0.543 \\
    MKW3s &  1.37 &  2.732 &   0.0444 & WINGSJ152151.84+074232.1 & cD &  -23.36 &  -25.61 &  48.87 &   5.9 &   0.691 &   0.600 &   0.590 &   0.590 &   0.557 &   0.548 \\
  RXJ0058 &  0.22 &  2.804 &   0.0484 & WINGSJ005822.88+265152.6 & cD &  -23.78 &         &  47.13 &   5.4 &   1.013 &   0.830 &   0.970 &   0.890 &   0.898 &   0.000 \\
  RXJ1022 &  0.18 &  2.761 &   0.0548 & WINGSJ102237.40+383445.0 &  E &  -22.33 &  -25.81 &  20.23 &   4.4 &   0.724 &   0.740 &   0.620 &   0.650 &   0.634 &   0.578 \\
  RXJ1740 &  0.26 &  2.765 &   0.0441 & WINGSJ174032.06+353846.1 & cD &  -22.75 &  -25.78 &  22.33 &   3.4 &   0.627 &   0.660 &   0.530 &   0.510 &   0.502 &   0.000 \\
 ZwCl1261 &  0.41 &        &   0.0644 & WINGSJ071641.24+532309.4 & cD &  -23.78 &  -27.08 &  43.48 &   4.8 &   0.711 &   0.730 &   0.610 &   0.480 &   0.530 &   0.000 \\
 ZwCl2844 &  0.29 &  2.729 &   0.0503 & WINGSJ100236.54+324224.3 & cD &  -23.21 &  -26.26 &  19.15 &   4.3 &   0.755 &   0.760 &   0.510 &   0.390 &   0.384 &   0.537 \\
 ZwCl8338 &  0.40 &  2.852 &   0.0494 & WINGSJ181105.18+495433.7 & cD &  -23.54 &  -26.59 &  25.45 &   4.9 &   0.845 &   0.840 &   0.800 &   0.810 &   0.801 &   0.692 \\
 ZwCl8852 &  0.48 &  2.884 &   0.0408 & WINGSJ231042.27+073403.7 &  E &  -23.26 &  -26.44 &  17.39 &   4.9 &   0.911 &   0.880 &   0.900 &   0.890 &   0.862 &   0.891 \\
 \hline
\end{tabular}
\end{minipage}
(a) Decimal logarithm of the X-ray luminosities in the 0.1-2.4~keV band from \citet{ebel96,ebel98,ebel00}\\
(b)-(c) Decimal logarithm of the velocity dispersions of cluster galaxies (b) and average redshifts of clusters (c) from \citet{cava09}. Some values have been updated, according to \citet{cava10}.\\
(d) WINGS identifier from \citet{vare09}. \\
(e) Morphological type (see text). \\
(f) Total absolute magnitude in the V-band from \citet[][corrected for galactic extiction]{vare09}\\
(g) Total absolute magnitude in the K-band from 2MASS.\\
(h) Effective radius in kpc from GASPHOT. Missing values indicate that GASPHOT failed to converge.\\
(i) Sersic index from GASPHOT. Missing values indicate that GASPHOT failed to converge.\\
(j) Axial ratio at the effective isophote from GASPHOT. In case GASPHOT failed to converge, we derive $q_{_G}$ from the equation: $q_{_G}$=0.392+0.572$\times q_{_S}$ (obtained from the linear best-fitting of the relation in panel 15 of Figure~\ref{axratcomp}.\\
(k)-(m) Axial ratios of the isophotes at 15, 30 and 60 kpc of major axis [(k), (l) and (m), respectively] from GASPHOT.\\
(n) Axial ratio at the threshold area from \citet[][SExtractor]{vare09}.\\
(o) Axial ratio from the LEDA Hypercat database \citep{patu03}.
\end{table}
\end{landscape}

\begin{landscape}
\begin{table}
\begin{minipage}{250mm}
\caption{Main properties of the additional cD sample}
\label{tab2}

\begin{tabular}{ccccccccccccc}
\hline\hline
Cluster & WINGS\_ID & rank$^{(a)}$ & 
$M_V$ & $M_K$ & R$_e$ & $n$ & 
$q_{_G}$ & $q_{_{15}}$ & $q_{_{30}}$ & 
$q_{_{60}}$ & $q_{_S}$ & $q_{_L}$ \\
 & & & & kpc & & & & & & & \\
\hline
    A548b & WINGSJ054527.59-255550.9 & 3 & -22.17 &  -25.00 &        &       &   0.740 &   0.750 &   0.790 &   0.790 &   0.608 &   0.000 \\
     A602 & WINGSJ075316.63+292405.3 & 2 & -21.87 &  -25.07 &        &       &   0.724 &   0.660 &   0.590 &   0.600 &   0.580 &   0.000 \\
    A1736 & WINGSJ132727.99-271929.2 & 2 & -23.76 &  -26.65 &  31.33 &   5.2 &   0.607 &   0.620 &   0.510 &   0.570 &   0.614 &   0.542 \\
    A1736 & WINGSJ132800.12-272115.5 & 3 & -23.15 &  -26.48 &        &       &   0.854 &   0.730 &   0.760 &   0.760 &   0.808 &   0.724 \\
    A1831 & WINGSJ135908.75+280121.3 & 2 & -23.36 &  -26.29 &        &       &   0.594 &   0.480 &   0.410 &   0.370 &   0.353 &   0.000 \\
    A1983 & WINGSJ145243.26+165413.5 & 2 & -22.54 &  -25.55 &   9.47 &   4.5 &   0.693 &   0.660 &   0.600 &   0.600 &   0.588 &   0.689 \\
    A3158 & WINGSJ034252.97-533752.6 & 3 & -22.64 &  -26.60 &  12.78 &   4.3 &   0.826 &   0.830 &   0.650 &         &   0.762 &   0.525 \\
    A3266 & WINGSJ043021.97-613200.7 & 2 & -22.87 &  -26.12 &        &       &   0.886 &   0.880 &   0.900 &   0.890 &   0.863 &   0.000 \\
    A3395 & WINGSJ062649.57-543234.5 & 2 & -24.09 &  -26.03 &        &       &   0.975 &   0.930 &   0.910 &   0.920 &   0.835 &   0.813 \\
   A3528a & WINGSJ125452.41-291617.1 & 2 & -22.30 &  -25.77 &        &       &   0.711 &   0.550 &   0.550 &   0.550 &   0.557 &   0.000 \\
    A3556 & WINGSJ132329.02-315039.6 & 2 & -24.41 &  -26.34 &        &       &   0.877 &   0.830 &   0.710 &   0.670 &   0.661 &   0.631 \\
 ZwCl8338 & WINGSJ181109.74+495153.0 & 2 & -22.47 &  -25.70 &        &       &   0.712 &   0.640 &   0.620 &   0.580 &   0.560 &   0.000 \\
 ZwCl8852 & WINGSJ231022.37+073450.5 & 2 & -22.84 &  -26.19 &  14.58 &   2.8 &   0.683 &   0.620 &   0.560 &   0.550 &   0.509 &   0.558 \\
 ZwCl8852 & WINGSJ231030.43+073520.6 & 3 & -22.89 &  -25.88 &        &       &   0.760 &   0.830 &   0.690 &   0.700 &   0.644 &   1.000 \\
 \hline
\end{tabular}
\end{minipage}
(a) ~Luminosity ranking in the V-band
\end{table}
\end{landscape}

\appendix

\section{Disentangling cDs from BCG\_Es with MORPHOT} \label{appmorph}

MORPHOT is an automatic tool for galaxy morphology, puposely devised
in the framework of the WINGS project. An exhaustive description of
the tool will be given in a forthcoming paper (Fasano et al., in
preparation). Here we just outline the logical sequence and the basic
procedures of MORPHOT. It extends the classical CAS
(Concentration/Asymmetry/clumpinesS) parameter set \citep{cons03}, by
using 20 image-based morphological diagnostics. Fourteen of them have
never been used, while the remaining six are actually already present
in the literature, although in slightly different forms: the Sersic
index, the Concentration index \citep{cons03}, the Gini and M20 
coefficients \citep{lotz04}, the Asymmetry and Clumpiness parameters
\citep{cons03}. 

Besides depending on the visual morphology, all diagnostic are also
empirically found to depend on the relative size (area enclosing 80\%
of the total light divided by FWHM area) and flattening of galaxies,
as well as on the image S/N ratio. These `a priori' dependencies have
been removed using a sizeable set of simulated galaxies and a control
sample of $\sim$1500 WINGS$+$SDSS galaxies, visually classified by two
of us (GF and AD). These visual classifications have been used to
calibrate and combine the 20 diagnostics, thus obtaining a final,
single estimator of the morphological type, equipped with the proper
confidence interval. This has been achieved by averaging the results
of two different, totally independent approaches, based on the Neural
Network (NN) and Maximum Likelihood (ML) techniques. 
The first approach (NN) is based on a particular kind of feed-forward 
Neural Network, called Multilayer Perceptron Artificial NN 
\citep[MLP, see][]{vanz04}. In this case the control galaxy sample provides the 
training set of the NN machine, with the morphological diagnostics 
used as inputs and the visual classifications as targets.
In the ML-based approach the probability density distributions of all
diagnostics for all morphological types are drawn from the control
galaxy sample. In this case the `blind' morphological type estimate of
any other galaxy is obtained maximizing the product of the
probabilities of the particular set of diagnostics derived from the
galaxy image.

\begin{figure}
\includegraphics[scale=0.45]{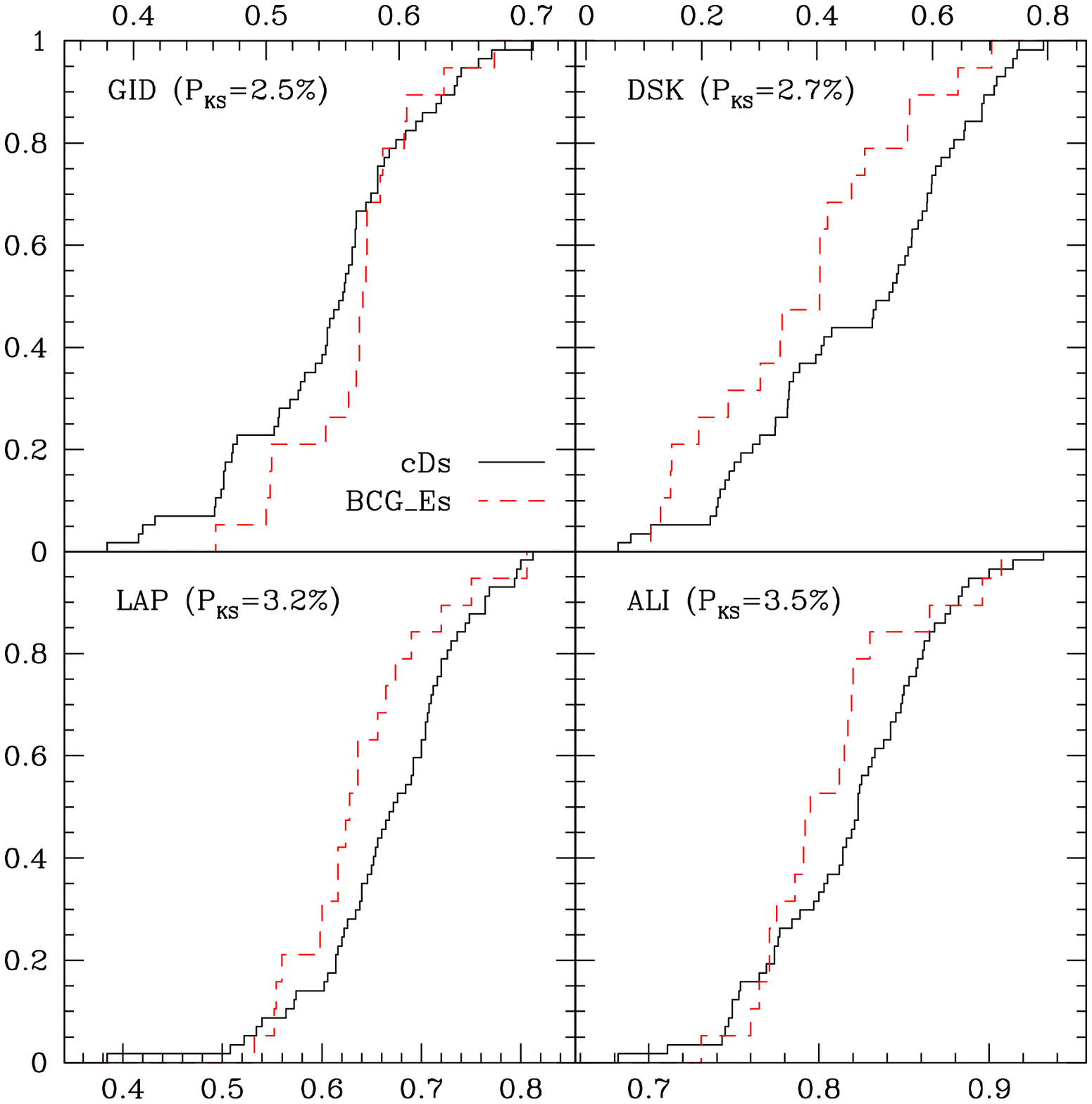}
\caption{
Comparison between cDs (full line) and BCG\_Es (dashed line; red in 
the electronic version) for the cumulative distributions 
of the morphological diagnostics $GID$, $DSK$, $LAP$ and $ALI$ (see 
text for a short description of them).
\label{morphot}}
\end{figure}

Defining each diagnostic and explaining its meaning is beyond the
scope of the present paper. Here we just mention that all diagnostics
are normalized to vary in the range (0--1) and that four
out of the 14 newly devised diagnostics turned out to be the most
effective ones in order to disentangle cDs from BCG\_Es. They are: 
(i) a modified version of the Gini coefficient ($GID$), where the
pixels are sorted according to the distance from the galaxy center,
rather than according to the flux \citep[as in][]{lotz04};
(ii) a Diskiness coefficient ($DSK$), measuring the correlation
between azimuth and pixel flux relative to the average flux value of
the elliptical isophote passing through the pixel itself;
(iii) an image-averaged 2D-Laplacian ($LAP$) in polar coordinates,
roughly quantifying the global degree of concaveness of the galaxy
profile;
(iv) an Alignment coefficient ($ALI$), measuring the average
concordance between the (local) maximum flux gradient direction and
the (local) direction of the galaxy center.

In Figure~\ref{morphot}, the cumulative distributions of the four
above mentioned diagnostics for cDs and BCG\_Es are reported for
comparison, together with the relative KS probabilities. The figure
shows that, compared with BCG\_Es, the cD galaxies have (on average)
lower values of $GID$ (less peaked profiles) and larger values of
$DSK$ (more disky), $LAP$ (more concave) and $ALI$ (local gradient
better aligned with the galaxy center). Even if none of the four
diagnostics illustrated in the figure, alone, turns out to be
conclusive, their combination (through MORPHOT) provides us with a
powerful tool for the particular task of disentangling cDs from
BCG\_Es.


\begin{thebibliography}{99}
\bibitem[\protect\citeauthoryear{Allgood et al}{2006}]{allg06} Allgood, B., Flores, R. A., Primack, J. R., Kravtsov, A. V., Wechsler, R. H., Faltenbacher, A., \& Bullock, J. S.  2006, MNRAS, 367, 1781
\bibitem[Bailin and Steinmetz(2005)]{bast05} Bailin, J., \& Steinmetz, M.  2005, ApJ, 627, 647
\bibitem[Bak and Statler(2000)]{bast00} Bak, J., \& Statler, T. S.  2000, AJ, 120, 110
\bibitem[Beers and Geller(1983)]{bege83} Beers, T. C., \& Geller, M. J.  1983, ApJ, 274, 491
\bibitem[Bernardi et al.(2007)]{bern07} Bernardi M., Hyde J. B., Sheth R. K., Miller C. J., \& Nichol R. C. 2007, AJ, 133, 1741
\bibitem[Bernardi(2009)]{bern09} Bernardi M. 2009, MNRAS, 395, 1491
\bibitem[Bertin and Arnouts(1996)]{bear96} Bertin, E., \& Arnouts, S.  1996, A\&A, 117, 393
\bibitem[Bett et al.(2007)]{bett07} Bett, P., Eke, V., Frenk, C. S., Jenkins, A., Helly, J., \& Navarro, J.  2007, MNRAS, 376, 215
\bibitem[Bildfell et al.(2008)]{bild08} Bildfell, C., Hoekstra, H., Babul, A., \& Mahdavi, A. 2008, MNRAS, 389, 1637
\bibitem[Bower et al.(2006)]{bowe06} Bower, R. G., Benson, A. J., Malbon, R., Helly, J. C., Frenk, C. S., Baugh, C. M., Cole, S., \& Lacey, C. G. 2006, MNRAS, 370, 645
\bibitem[Brough et al.(2002)]{brou02} Brough, S., Collins, C. A., Burke, D. J., Mann, R. G., \& Lynam, P. D. 2002, MNRAS, 329, L53
\bibitem[Brough et al.(2005)]{brou05} Brough, S., Collins, C. A., Burke, D. J., Lynam, P. D., \& Mann, R. G. 2005, MNRAS, 364, 1354
\bibitem[Burke et al.(2000)]{burk00} Burke, D. J., Collins, C. A., \& Mann, R. G. 2000, ApJLett., 532, 105
\bibitem[Carter and Metcalfe(1980)]{came80} Carter, D. \& Metcalfe, N.  1980, MNRAS, 191, 325
\bibitem[Cava et al.(2009)]{cava09} Cava, A., et al.  2009, A\&A, 495, 707
\bibitem[Cava et al.(2010)]{cava10} Cava, A., et al.  2010, in preparation
\bibitem[Cole and Lacey(1996)]{cola96} Cole, S., \& Lacey, C.  1996, MNRAS, 281, 716
\bibitem[Collins et al.(2009)]{coll09} Collins, C. A., et al. 2009, nature, 458, 603
\bibitem[Conselice(2003)]{cons03} Conselice, C. J., 2003, ApJS, 147, 1
\bibitem[Coziol et al.(2009)]{cozi09} Coziol, R., Andernach, H., Caretta, C. A., Alamo-Martínez, K. A., \& Tago, E. 2009, AJ, 137, 4795
\bibitem[De Lucia and Blaizot(2007)]{delu07} De Lucia G., \& Blaizot J. 2007, MNRAS, 375, 2
\bibitem[Desroches et al.(2007)]{desr07} Desroches, L. B., Quataert, E. Ma, C. P., \& West, A. A. 2007, MNRAS, 377, 402
\bibitem[D'Onofrio et al.(2009)]{dono09} D'Onofrio, M., et al. 2009, in preparation
\bibitem[Dubinski(1998)]{dubi98} Dubinski, J.  1998, ApJ, 502, 141
\bibitem[Ebeling et al.(1996)]{ebel96} Ebeling, H., Voges, W., Bohringer, H., Edge, A. C., Huchra, J. P., \& Briel, U. G.  1996, MNRAS, 281, 799
\bibitem[Ebeling et al.(1998)]{ebel98} Ebeling, H., Edge, A. C., Bohringer, H., Allen, S. W., Crawford, C. S., Fabian, A. C., Voges, W., \& Huchra, J. P.  1998, MNRAS, 301, 881
\bibitem[Ebeling et al.(2000)]{ebel00} Ebeling, H., Edge, A. C., Allen, S. W., Crawford, C. S., Fabian, A. C., \& Huchra, J. P.  2000, MNRAS, 318, 333
\bibitem[Edge and Stewart(1991)]{edst91} Edge, A. C., \& Stewart, G. C.  1991, MNRAS, 252, 428
\bibitem[Eke et al.(1996)]{eke96} Eke, V. R., Cole, S., Frenk, C. S., \& Navarro, J. F. 1996, MNRAS, 282, 263
\bibitem[Fabian(1994)]{fabi94} Fabian, A. C.  1994, Ann.Rev.Astron. \& Astrphys., 32, 277
\bibitem[Fasano and Vio(1991)]{favi91} Fasano, G., \& Vio, R.  1991, MNRAS, 249, 629
\bibitem[Fasano et al.(1993)]{fasa93} Fasano, G., Amico, P., Bertola, F., Vio, R., \& Zeilinger, W. W.  1993, MNRAS, 262,109
\bibitem[Fasano et al.(2006)]{fasa06} Fasano, G., et al.  2006, A\&A, 445, 805
\bibitem[Fasano et al.(2007)]{fasa07} Fasano, G., et al.  2007, From Stars to Galaxies: Building the Pieces to Build Up the Universe, A. Vallenari, R. Tantalo, L. Portinari, \& A. Moretti, ASP Conference Series, 374, 495
\bibitem[Flores et al.(2007)]{flores07} Flores, R. A., Allgood, B., Kravtsov, A. V., Primack, J. R., Buote, D. A., \& Bullock, J. S. 2007, MNRAS, 377, 883
\bibitem[Franx et al.(1991)]{fran91} Franx, M., Illingworth, G., \& de Zeeuw, T.  1991, ApJ, 383, 112
\bibitem[Gallagher and Ostriker(1972)]{gaos72} Gallagher, J. S. III, \& Ostriker, J. P.  1972, AJ, 77, 288
\bibitem[Gao et al.(2004)]{gao04} Gao, L., White, S.D.M., Jenkins, A., Stoehr, F., \& Springel, V. 2004, MNRAS, 355, 819
\bibitem[Giacintucci et al.(2007)]{giac07} Giacintucci, S., Venturi, T., Murgia, M., Dallacasa, D., Athreya, R., Bardelli, S., Mazzotta, P., \& Saikia, D. J.  2007, A\&A, 476, 99
\bibitem[Gottl\"ober and Yepes(2007)]{gott07} Gottl\"ober, S., \& Yepes, G. 2007, ApJ, 664, 117
\bibitem[Hashimoto et al.(2008)]{has08} Hashimoto, Y., Henry, J. P., \& Boehringer, H.  2008, MNRAS, 390, 1562
\bibitem[Hoessel et al.(1987)]{hoes87} Hoessel, J. G., Oegerle, W. R., \& Schneider, D. P.  1987, AJ, 94, 1111
\bibitem[Jones and Forman(1984)]{jofo84} Jones, C., \& Forman, W.  1984, ApJ, 276, 38
\bibitem[Katayama et al.(2003)]{kata03} Katayama, H., Hayashida, K., Takahara, F., \& Fujita, Y.  2003, ApJ, 585, 687
\bibitem[Kazantzidis et al.(2004)]{kaza04} Kazantzidis, S., Kravtsov, A. V., Zentner, A. R., Allgood, B., Nagai, D., \& Moore, B. 2004, ApJLett, 611, L73
\bibitem[Kimm and Yi(2007)]{kiyi07} Kimm, T., \& Yi, S. K.  2007, ApJ, 670, 1048
\bibitem[Kormendy and Djorgovski(1989)]{kodj89} Kormendy, J., \& Djorgovski, S.  1989, Ann. Rev. Astron. \& Astrophys., 27, 235
\bibitem[Laine et al.(2003)]{lain03} Laine, S., van der Marel, R. P., Lauer, T. R., Postman, M., O'Dea, C. P., \& Owen, F. N.  2003, AJj, 125, 478
\bibitem[Lin and Mohr(2004)]{lin04} Lin, Y., \& Mohr, J. J. 2004, ApJ, 617, 879
\bibitem[Lotz et al.(2004)]{lotz04} Lotz, J. M., Primack, J., \& Madau, P. 2004, AJ, 128, 163
\bibitem[Loubser et al.(2009)]{loub09} Loubser, S. I., S\'anchez-Bl\'azquez, P., Sansom, A. E., \& Soechting, I. K. 2009, MNRAS, 398, 133
\bibitem[Lucy(1974)]{lucy74} Lucy, L. B.  1974, AJ, 79, 745
\bibitem[Macci\`o et al.(2008)]{maccio08} Macci\`o, A. V., Dutton, A. A., \& van den Bosch, F. C. 2008, MNRAS 391, 1940
\bibitem[M\'endez-Abreu et al.(2008)]{mend08} M\'endez-Abreu, J., Aguerri, J. A. L., Corsini, E. M. \& Simonneau, E. 2008, A\&A, 478, 353
\bibitem[Merritt(1985)]{merr85} Merritt, D.  1985, ApJ, 289, 18
\bibitem[Noerdlinger(1979)]{noer79} Noerdlinger, P. D.  1979, ApJ, 234, 802
\bibitem[Oegerle and Hoessel(1991)]{oeho91} Oegerle, W. R., \& Hoessel, J. G.  1991, ApJ, 375, 15
\bibitem[Oemler(1976)]{oeml76} Oemler, A. Jr.  1976, ApJ, 209, 693
\bibitem[Ostriker and Tremaine(1975)]{ostr75} Ostriker, J. P., \& Tremaine, S. D.  1975, ApJLett., 202, 1130
\bibitem[Patel et al.(2006)]{pate06} Patel, P.; Maddox, S., Pearce, F. R., Arag\`on-Salamanca, A., \& Conway, E.  2006, MNRAS, 370, 851
\bibitem[Paturel et al.(2003)]{patu03} Paturel, G., Petit, C., Prugniel, Ph., Theureau, G., Rousseau, J., Brouty, M., Dubois, P., \& Cambr\`esy, L.  1993, A\&A, 412, 45
\bibitem[Pignatelli et al.(2006)]{pign06} Pignatelli, E., Fasano, G., \& Cassata, P.  2006, A\&A, 446, 373
\bibitem[Plionis et al.(2003)]{plio03} Plionis, M., Benoist, C., Maurogordato, S., Ferrari, C., \& Basilakos, S.  2003, ApJ, 594, 144
\bibitem[Poggianti et al.(2009)]{pogg09} Poggianti, B. M., et al.  2009, ApJLett., 697, 137
\bibitem[Richstone(1975)]{rich75} Richstone, D. O.  1975, ApJ, 200, 535
\bibitem[Richstone(1976)]{rich76} Richstone, D. O.  1976, ApJ, 204, 642
\bibitem[Ryden(1992)]{ryde92} Ryden, B.  1992, ApJ, 396, 445
\bibitem[Ryden et al.(1993)]{rlp93} Ryden, B. S., Lauer, T. R., \& Postman, M.  1993, ApJ, 410, 515
\bibitem[Sandage and Hardy(1973)]{sanh73} Sandage, A., \& Hardy, E.  1973, ApJSupp., 183, 743
\bibitem[Schneider et al.(1983)]{schn83} Schneider, D. P., Gunn, J. E., \& Hoessel, J. G.  1983, ApJ, 268, 476
\bibitem[Schombert(1986)]{sch86} Schombert, J. M.  1986, ApJSupp, 60, 603
\bibitem[Schombert(1987)]{scho87} Schombert, J. M.  1987, ApJSupp., 64, 643
\bibitem[Schombert(1988)]{scho88} Schombert, J. M.  1988, ApJ, 328, 475
\bibitem[Schombert(1992)]{sch92} Schombert, J. M.  1992, Morphological and Physical Classification of Galaxies, G. Longo, M. Capaccioli, \& G. Busarello, A\&A, 178, 53
\bibitem[Silk(1976)]{silk76} Silk, J.  1976, ApJ, 208, 646
\bibitem[Stott et al.(2008)]{stot08} Stott, J. P., Edge, A. C., Smith, G. P., Swinbank, A. M. \& Ebeling, H. 2008, MNRAS, 384, 1502
\bibitem[Struble(1990)]{str90} Struble, M. F.  1990, AJ, 99, 743
\bibitem[Thuan and Romanishin(1981)]{thro81} Thuan, T. X., \& Romanishin, W.  1981, ApJ, 248, 439
\bibitem[Tremblay and Merritt (1996)]{trem96} Tremblay, B., \& Merritt, D.  1996, AJ, 111, 2243
\bibitem[Valentinuzzi et al.(2009)]{vale09} Valentinuzzi, T., et al.  2009, [arXiv:0907.2392]
\bibitem[Varela et al.(2009)]{vare09} Varela, J., et al.  2009, A\&A, 497, 667
\bibitem[Vanzella et al.(2004)]{vanz04} Vanzella, E., Cristiani, S., Fontana, A., Nonino, M., Arnouts, S., Giallongo, E., Grazian, A., Fasano, G., Popesso, P., Saracco, P., \& Zaggia, S.  2004, A\&A, 423, 761
\bibitem[von~der~Linden et al.(2007)]{vond07} von~der~Linden, A., Best, P. N., Kauffmann, G., White, S. D. M.  2007, MNRAS, 379, 867
\bibitem[Wang et al.(2008)]{wang08} Wang, Y., Yang, X., Mo, H. J., Li, C., van den Bosch, F. C., Fan, Z., \& Chen, X. 2008, MNRAS, 385, 1511
\bibitem[Warren et al.(1992)]{warr92} Warren, M. S., Quinn, P. J., Salmon, J. K., \& Zurek, W. H.  1992, ApJ, 399, 405
\bibitem[West(1994)]{west94} West, M. J.  1994, MNRAS, 268, 79
\bibitem[Whiley et al.(2008)]{whil08} Whiley, I. M., et al. 2008, MNRAS, 387, 1253
\bibitem[White(1976)]{whit76} White, S. D. M.  1976, MNRAS, 177, 717

\end{thebibliography}
\end{document}